\documentclass[manuscript,11pt]{aastex}

\usepackage{graphicx}
\usepackage{enumerate}

\input blackdvi.tex

\lefthead{Levison et al$.$}
\righthead{Giant Planet Core Formation}

\begin{document}

\title{Modeling the Formation of Giant Planet Cores I: Evaluating Key
Processes}

\author{Harold F$.$ Levison}
\affil{Southwest Research Institute\\ 
1050 Walnut St, Suite 300\\ 
Boulder, CO 80302}
\authoremail{hal@boulder.swri.edu}

\author{Edward Thommes}
\affil{Department of Physics\\ University of Guelph\\ Guelph, Ontario,
N1G 2W1\\ Canada}

\and

\author{Martin J$.$ Duncan}
\affil{Department of Physics\\ Queen's University\\
Kingston, Ontario, K7L 3N6\\ Canada}


\received{\Red{July 15, 2009}}

\accepted{December 9, 2009}

\newpage

\begin{abstract}
  
  One of the most challenging problems we face in our understanding of
  planet formation is how Jupiter and Saturn could have formed before
  the the solar nebula dispersed. The most popular model of giant
  planet formation is the so-called {\it core accretion} model. In
  this model a large planetary embryo formed first, mainly by two-body
  accretion. This is then followed by a period of inflow of nebular
  gas directly onto the growing planet (Pollack et al$.$~1996). The
  core accretion model has an Achilles heel, namely the very first
  step. We have undertaken the most comprehensive study of this
  process to date.  In this study we numerically integrate the orbits
  of a number of planetary embryos embedded in a swarm of
  planetesimals.  In these experiments we have included a large number
  of physical processes that might enhance accretion. In particular,
  we have included: 1) aerodynamic gas drag, 2) collisional damping
  between planetesimals, 3) enhanced embryo cross-sections due to
  their atmospheres, 4) planetesimal fragmentation, and 5)
  planetesimal driven migration.  We find that the gravitational
  interaction between the embryos and the planetesimals lead to the
  wholesale redistribution of material --- regions are cleared of
  material and gaps open near the embryos. Indeed, in 90\% of our
  simulations without fragmentation, the region near that embryos is
  cleared of planetesimals before much growth can occur. Thus, the
  widely used assumption that the surface density distribution of
  planetesimals is smooth can lead to misleading results.  In the
  remaining 10\% of our simulations, the embryos undergo a burst of
  outward migration that significantly increases growth. On timescales
  of $\sim\!10^5$ years, the outer embryo can migrate $\sim\!6\,$AU
  and grow to roughly $30\,M_\oplus$. This represents a largely
  unexplored mode of core formation.  We also find that the inclusion
  of planetesimal fragmentation tends to inhibit growth except for a
  narrow range of fragment migration rates.

\end{abstract}


\newpage
\section{Introduction}
\label{sec:intro}

It is ironic that the most massive planets in the Solar System had to
have formed in the least amount of time. Jupiter and Saturn, for
example, which are made mainly of hydrogen and helium, must have
accreted this gas before the solar nebula dispersed. Observations of
young star systems (e$.$g$.$, Haisch, Lada \& Lada 2001\Red{; see
  Hillenbrand~2008 and references therein for a recent review}) show
that gas disks, at least insofar as they are traced by the presence of
dust in the inner AU as well as accretion onto the star, have
lifetimes of $\sim\!1-10\,$Myr. So, the gas giant planets had to form
before this time. In contrast, the Earth most likely took at least
$60\,$Myr to fully form (based on cosmochemical constraints;
Halliday~2004, Touboul et al$.$~2007; and numerical modeling; Chambers
\& Wetherill~1998; Agnor, Canup, \& Levison~1999), and may have taken
as long as $100\,$Myr.

Thus, one of the most challenging problems we face in our
understanding of planet formation is how Jupiter and Saturn could have
formed so quickly. In the core accretion model, which envisions that
a large planetary embryo formed first by two-body accretion followed
by a period of inflow of gas directly onto the growing planet (Mizuno
et al$.$~1978; Pollack et al$.$~1996), the main difficulty is in the
first step. The accretion of a massive atmosphere requires a solid
core $\sim\!10\,M_\oplus$ in mass (Mizuno~1980; Pollack et al$.$~1996;
Hubickyj et al$.$~2005). As we describe in detail below, assembling
such a large body, it turns out, offers some serious challenges to the
theory of planet formation as it currently stands. The difficulties
are threefold: First, the accretion process has to be efficient enough
to concentrate such a large mass in (at least) one single body.
Second, everything has to happen fast enough ($\lesssim\!10^7\,$yr,
for reasons described above) that when the putative core is ready,
there is still enough gas---of order several hundred $M_\oplus$---left
in the nearby part of the disk to furnish its envelope. The final
problem concerns migration due to planet-disk tidal interactions,
which threatens to drop core-sized bodies into the central star faster
than they can accrete (Ward~1986; Korycansky \& Pollack~1993;
Ward~1997).

In the last five years or so, there has been a concerted effort by the
planet formation community to overcome these problems. Indeed,
several new ideas have been presented in the literature (see the
review in \S{\ref{sec:review}} below). However, many of these ideas
have yet to be fully explored with modern dynamical simulations.
Thus, here we present a series of direct numerical $N$-body
simulations intended to explore these ideas. In particular, we employ
a simplistic set of initial conditions that are designed to study the
effectiveness of various physical processes rather than realistically
model the growth of giant planet cores. This paper is organized as
follows. After our review in \S{\ref{sec:review}}, we describe our
numerical methods in \S{\ref{sec:code}}, our results in
\S\ref{sec:results}, and conclude in \S{\ref{sec:concl}.

\section{A Not-So Brief Review of Core Accretion Models}
\label{sec:review}

Perhaps the best known simulations of the core accretion scenario are
those of Pollack et al$.$ (1996, see Hubickyj, et al$.$~2005 for an
updated version of these models). These simulations follow the growth
of a single isolated embryo as it first accreted neighboring
planetesimals, and then nebular gas. Although these models show that
\Red{there are reasonable conditions under which it is} possible for
Jupiter and Saturn to form in less than $10\,$Myr, they employed a
simplistic model of solid body accretion. Indeed, these models are
1-dimensional and only mimic the dynamical evolution of the system
using crude expressions for dynamical stirring and the evolution of
the surface density of the planetesimals. As we now describe, more
realistic dynamical models fail, in general, to form cores large
enough to undergo significant gas accretion before the nebula
disperses.

Studies of terrestrial planet formation have shown that solid body
growth can occur in three stages. In the first stage, planetesimals
grow by runaway accretion, wherein the largest bodies grow the fastest
(Wetherill \& Stewart~1989). For the minimum-mass solar nebula
(hereafter MMSN, Hayashi~1981) Ida \& Makino (1993) argued that
runaway accretion stops at a protoplanet mass of only
$\sim\!10^{-6}M_\oplus$. 

In the middle stage, \Red{known as the `{\it oligarchic regime}',}
accretion changes from runaway to self-regulating, as the largest
bodies become big enough to gravitationally ``stir their own soup'' of
planetesimals (Ida \& Makino 1993, Kokubo \& Ida 1998, 2000, Thommes,
Duncan \& Levison 2003, hereafter TDL03). In this phase, the largest
few objects at any given time are of comparable mass, and are
separated by amounts determined by their masses and distances from the
Sun. As the system evolves, the mass of the system is concentrated
into an ever-decreasing number of bodies of increasing masses and
separations.  This stage ends at a given location in the disk when the
local ``oligarchy'' of largest bodies reach their {\it isolation mass}
(at least within a factor $\sim\!2$), meaning that they have consumed
all planetesimals within their gravitational reach. In the terrestrial
planet region, typical disk models produce isolation masses of only
about Mars mass, thus a third \Red{stage} must take place in which
these bodies' orbits cross and they collide to form Earth- and
Venus-mass bodies.

In the Jupiter-Saturn region, our current understanding is that the
first and second stages occurred, but it is possible that the final
stage did not. There are two arguments for this. First, in the
terrestrial region the last stage of planet formation takes 30 --
$50\,$Myr years in the standard gas-free model (Chambers \& Wetherill
1998; Agnor, Canup, \& Levison~1999; Chambers 2001), which implies
that it should take at least an order of magnitude longer in the
Jupiter-Saturn region because of lower densities and longer orbital
periods. This is much longer than the lifetime of the gas disks.
Second, while the isolation masses in the terrestrial zone are much
smaller than the observed planets, in the Jupiter-Saturn region the
isolation mass can be roughly what is needed for gas accretion
($\sim\!10\,M_\oplus$), if one were to accept a significantly enhanced
disk mass (Lissauer 1987). Thus, oligarchic growth could, in
principle, suffice to produce the giant planet cores.

During oligarchic growth, the planetary embryos enhance their mass by
sweeping up much smaller planetesimals, although occasional mergers
between embryos do occur (see Thommes \& Duncan~2006, hereafter TD06,
for a review). The rate of growth is determined by the local surface
density of planetesimals, $\Sigma_m$, and the random velocities of
both the embryos and the planetesimals. The embryos are self-excited,
while they are being damped by dynamical friction with the
planetesimals. \Red{Since during oligarchic growth the overall density
  of the embryo population is small compared to that of the
  planetesimal disk,} dynamical friction dominates and the embryos are
on circular, co-planar orbits. In this case, Kokubo \& Ida (1998) show
that the embryos are well separated from one another such that $\Delta
a = b\, r_H,$ where $ r_H = a \left( M/3 M_\odot\right )^{1/3}$ is the
Hill radius of the protoplanet, $a$ is heliocentric distance, and $b$
is a constant with a value of $\sim5-10$.

The planetesimals' random velocities in the oligarchic regime are
dominated by stirring due to the protoplanets (indeed this is how this
regime is defined). In most of the work that has been done (see below
for more detail), it is assumed that this is balanced by the
aerodynamic drag due to the nebular gas. One can estimate the
resulting equilibrium RMS eccentricity by equating the timescales for
these two effects. It is found that $e_m \propto M^{1/3} r^{1/5}
\rho_{\rm gas}^{-1/5}$, where $\rho_{\rm gas}$ is the local density of
the nebula, and $r$ is the radius of a typical planetesimal. Note
that the embryo growth rate is $\propto \Sigma_m e_m^{-2} \propto
\Sigma_m r^{-2/5} \rho_{\rm gas}^{2/5}$, and thus the dynamically {\it
colder} the system of planetesimals (i.e$.$ the smaller $e_m$), the
faster the embryos grow. Keeping the system cold is key to growing
the cores before the nebular gas disappears. This can only happen if
the drag forces are large, which, in turn, implies that $\rho_{\rm
gas}$ needs to be large and $r$ small.

But, there is a price to be paid for large drag forces. In addition
to damping eccentricities, aerodynamic gas drag extracts energy from
planetesimal orbits (Adachi et al$.$~1976). Indeed, with the balance
between damping and gravitational stirring by the protoplanets
maintaining a nonzero equilibrium planetesimal random velocity, there
is a continuous net orbital decay of planetesimals. The surface
density of planetesimals thus changes at a given radius not just
because planetesimals are swept up by protoplanets, but also because
of this migration. This can have a serious negative effect on the
growth of the embryos.

TDL03 studied this in detail using both analytical analysis and
numerical $N$-body experiments. They find that gas drag acts as a
two-edged sword in the accretion of massive bodies: On the one hand,
increasing the strength of gas drag (by increasing the gas density,
decreasing the planetesimal size, or a combination thereof) damps
random velocities more strongly and speeds the accretion rate. On the
other hand, as also found by Inaba \& Wetherill (2001), stronger gas
drag also increases the rate at which $\Sigma_m$ is depleted by
planetesimal orbital decay; this causes growth to stall earlier than
we might otherwise expect. In particular, assuming a characteristic
planetesimal size of 10 km, TDL03 found that in order to produce
$10\,M_\oplus$ protoplanets in the Jupiter-Saturn region requires a
very massive disk ($\sim\!10$ times the MMSN). Such a large disk is
difficult to reconcile with other constraints, such as those derived
from the migration of Uranus and Neptune (Hahn \& Malhotra~1999;
Gomes, Morbidelli, \& Levison~2004). The good news is, however, that
the time to reach $10\,M_\oplus$ at $5\,$AU is short, roughly
$1\,$Myr. Decreasing the characteristic planetesimal size decreases
the mass at which growth stalls.

Unfortunately, there is more bad news: TDL03 did not include the
effects of fragmentation, which leads to considerably less optimistic
results. Fragmentation reprocesses a large fraction of the
planetesimals to much smaller size, which makes the above problem even
worse. The dynamical regime we are considering --- wherein the random
velocities of the planetesimals are determined by stirring from much
larger bodies --- makes it likely that fragmentation will play a role,
since planetesimals will collide with relative velocities much larger
than their surface escape velocities. Inaba \& Wetherill~(2001)
studied this effect using a statistical rather than an $N$-body code.
They found that the largest protoplanet produced has a mass of less
than $2\,M_\oplus$! This result is thus worrisome for the core
accretion model.

In recent years there have been several papers that suggest promising
avenues for solving weaknesses in the oligarchic growth model for
giant planet core accretion. They include:

\medskip

\begin{itemize}

\item{\underline{\it The role of embryo atmospheres:}} The model of
Pollack et al. (1996) for the accretion of giant planet cores also
models the slow accretion of a gas atmosphere onto a growing giant
embryo, prior to the final runaway gas accretion phase. This
atmosphere acts to enhance the capture cross-section for additional
planetesimals, but because they assume planetesimal sizes of
1-100$\,$km, the tenuous atmosphere present during the initial
solids-dominated accretion phase does not play a very large role in
raising the accretion rate. However, as a planetesimal's size is
decreased, the strength of gas drag it feels is increased. So too,
therefore, is its capture radius with respect to a embryo possessing
a gas atmosphere. Inaba \& Ikoma (2003), hereafter II03, study the
capture of planetesimals in the atmosphere of a growing core. Using
a simple 1-dimensional numerical model they show that, as long as
random velocities of the planetesimals are small compared to the
escape velocity from the core's surface, the effective cross-section
of the embryos is significantly increased. This can lead to a
substantial increase in accretion rate.

\item[] Using a Monte Carlo technique, Inaba, Wetherill \&
  Ikoma~(2003; hereafter IWI03) studied core formation with
  fragmentation, radial migration, and gas-enhanced capture
  cross-sections. They found that the inclusion of embryo atmospheres
  effectively rescues the core-accretion model from the perils of
  fragmentation. The important point is that fragmentation now has a
  positive as well as a negative consequence: Smaller planetesimals
  are lost more rapidly by migration, but are also accreted more
  readily by cores with atmospheres. Simply put, the two effects
  largely cancel each other out, and the results of IWI03 are not
  dissimilar from those of TDL03: For a disk of about ten times the
  solids and gas density of the minimum-mass model, a $\sim
  20\,M_\oplus$ body forms at 5 AU in $\sim\!3\times 10^6$ years.
  Unfortunately, even with this improvement, \Red{IWI03 finds that}
  very massive disks are required to build Jupiter's core before the
  gas disk dissipates, and finds that and it is still not possible to
  accrete Saturn's in time.

\item{\underline{\it Accretion in the shear-dominated regime:}} In all
the work discussed thus far, it was assumed that the velocity
dispersion of the planetesimals, $v_m$, was large enough that the
scale height of the planetesimal disk exceeds the radius of the
Hill's sphere, i.e$.$ the disk behaves as if it is fully three
dimensional. This occurs if $v_m \gtrsim \Omega r_H$, where
$\Omega$ is the orbital frequency of the embryo. This situation is
referred to as the {\it dispersion-dominated} regime. However, if
gas drag damps planetesimal random velocities strongly enough, $v_m$
can get much smaller than $\Omega r_H$ and the system enters the
so-called {\it shear-dominated regime}.

\item[] Rafikov (2004; hereafter R04) studied the above situation with
  an order-of-magnitude analytic analysis. He found that
  shear-dominated oligarchic growth proceeds in a qualitatively
  different way, and can be much more rapid, than dispersion-dominated
  growth (also see Goldreich et al$.$~2004a,b). The important
  distinction is that the damping of random velocities is so rapid
  that between consecutive close encounters with a growing embryo, a
  planetesimal loses almost all of its eccentricity and inclination.
  This makes the planetesimal disk very thin, which increases the
  accretion rate onto a planetary embryo. In the most extreme case,
  the velocity dispersion can be so small that the entire vertical
  column of the planetesimal disk has a scale height less than the
  protoplanet's Hill radius, thus making accretion a two-dimensional
  process. In this case, the accretion rate can be much larger than
  \Red{what might be expected in a purely three dimensional situation
    because planetesimals can no longer pass above or below the
    embryo.}  For example, using R04's analytic estimates, at 5 AU the
  two-dimensional accretion rate is roughly an order of magnitude
  larger than the oligarchic growth rate with a MMSN and 1 km-sized
  planetesimals.
  
\item[] \Red{At $5\,$AU, the transition between shear- and
    dispersion-dominated accretion occurs for planetesimals having
    sizes of roughly $\sim\!100\,$m - $1\,$km in a minimum mass solar
    nebula.  The} transition size increases with distance from the
  star. In reality, it is likely only a fraction of the total
  planetesimal population which finds itself in this regime because
  these objects have a broad size distribution. However, R04 shows
  that only 1\% of the total mass in planetesimals needs to be
  shear-dominated in order for the accretion rate of embedded
  protoplanets to be dominated by this part of the population.
  Indeed, R04 favors an idea similar to that of Wetherill \&
  Stewart~(1993), in which planetesimals are in the
  dispersion-dominated regime, but the small products of their mutual
  collisions are in the shear-dominated domain and are quickly
  accreted.

\item[] While the work in R04 may supply the long-sought-after
solution to giant planet core accretion, Rafikov was forced to make
a few significant simplifications in order to perform his analytic
analysis. For example, he ignored the role of planetesimal radial
migration. Just as a shear-dominated planetesimal receives discrete
kicks in random velocity which are then damped, so too will it
orbitally decay in discrete jumps, which will tend to remove
planetesimals from a protoplanet's feeding zone.

\item[] In addition, Rafikov was forced to assume that the surface
  density of the disk particles will remain spatially smooth. In
  reality, just as is seen in Saturn's rings, the planetary embryos
  will attempt to open gaps in a shear-dominated disk which will
  significantly lower the accretion rates. R04 suggests that the
  presence of other embryos might prevent gap opening. However, direct
  dynamical simulations of problems with similar dynamics have shown
  that this is not the case (McNeil, Duncan, \& Levison~2005; Levison
  \& Morbidelli 2007). In particular, gaps are opened in situations
  where the embryos are well separated, while the particles become
  trapped in the L$_4$ and L$_5$ Lagrange points if the embryos are
  closely packed. Thus, the assumption that the disk remains spatially
  smooth is probably not correct. However, this needs to be studied
  with realistic numerical experiments in order to determine whether
  gaps open in the case of giant planets cores, and if they do, how
  the accretion rate is affected. This is a major goal of this work.
  
\item{\underline{\it The role of collisional damping:}} A new scenario
  for the {\it in situ} formation of Uranus and Neptune has been
  proposed by Goldreich et al$.$~(2004a; 2004b, hereafter G04ab).
  While this new scheme was developed in a gas-free environment, there
  are elements that could be relevant to solving the core-formation
  timescale problem. In particular, G04ab envision that during the
  early stages of oligarchic growth, the growing embryos begin to
  dynamically excite the planetesimals. As the planetesimal population
  heats up, collisions between planetesimals become destructive. As a
  result, the number of planetesimals increase and thus the
  collisional mean free path decreases. The \Red{increased} collision
  rate, in turn, damps the planetesimal's random velocity.  So,
  independent of the initial sizes of the planetesimals, the system
  naturally evolves into the shear-dominated regime where accretion
  rates can be large.

\item[] The collisional damping mechanism has a distinct advantage
  over aerodynamic drag in that it does not cause radial migration.
  Thus, it might be possible to preserve $\Sigma_m$ while keeping
  $v_m$ small. However, this mechanism still potentially suffers from
  the gap-opening problem described above (Levison \& Morbidelli
  (2007)). Again, one of the goal of this work is to evaluate the
  importance of this problem.

\item{\underline{\it The role of evaporation/condensation:}} Jupiter
and Saturn formed in a region of the Solar System where water acts
like a solid. Water was so abundant in the solar nebula that the
surface density of solids was roughly a factor of 4 larger in
regions where water condensed than in areas where it did not.
Condensation occurred at the location known as the {\it snow-line}.
We might be able to use the proximity of the giant planets to the
snow-line to enhance the solid surface density to even higher levels
(Stevenson \& Lunine~1988; Cuzzi \& Zahnle~2004, hereafter CZ04).
From our perspective, the most interesting idea was put forward in
CZ04, as we now discuss.

\item[] One of the recurring themes in our discussions of core
accretion thus far is that the rapid inward migration of
planetesimals acts to frustrate planet growth. However, under the
right circumstances, CZ04 showed that the opposite can also be true.
Indeed, in their model, things actually work {\it best} if
planetesimals are of order meters rather than kilometers in size.
The resulting rapid radial migration serves as a means to produce a
high local concentration of condensible material, specifically ice.
This happens because when ice-rich planetesimals from the outer disk
arrive at the snow-line, their complement of water begins to
evaporate. CZ04 model the inward flux of ice-rich planetesimals, the
water evaporation rate and the diffusion rate of the vapor plume
which results, and show that a steady state may not exist until a
very large local enhancement in water, between 1 and 2 orders of
magnitude, has occurred. Since this vapor plume straddles the snow
line, water will recondense onto solid bodies at its outer edge. In
this way, the protoplanetary disk receives a large enhancement in
ice over a relatively small radial range ($\Delta a \lesssim 1$ AU).
Since accretion rate is $\propto \Sigma_{\rm solid}$, and isolation
mass is $\propto \Sigma_{\rm solid}^{3/2}$, an increase in the
surface density of solids makes it possible to grow bigger bodies
faster. Such an enhancement in $\Sigma_{\rm solid}$ has never been
included in a model of core accretion.

\end{itemize}

The only work of which we are aware that attempted to model most of
the above processes in a self consistent way is presented in
Chambers~(2008). Using a semi-analytic model, Chambers followed the
growth of planetary embryos in the giant planet region while they were
embedded in a swarm of planetesimals. He included the effects of
aerodynamic drag and fragmentation on the planetesimals, and
planet-disk tidal interactions (so-called Type~I migration; Ward~1986)
on the embryos. He found that under a reasonable set of initial
conditions, he can produce cores large enough to accrete gas within
the disk lifetime.

Although \Red{ambitious, Chambers}~(2008) used semi-analytic methods
to follow the evolution of the planetesimals, and thus did not fully
account for processes such as gap opening, which has been shown to
negatively effect accretion rates in the shear-dominated regime
(McNeil, Duncan, \& Levison~2005; Levison \& Morbidelli 2006). He also
did not include the effects of planetesimal driven planet migration
(Fernandez \& Ip~1984; Hahn \& Malhotra~1999; Gomes et al$.$~2003,
Kirsh et al$.$ 2009).

The effects of planetesimal-driven migration have largely been ignored
when gas is present because it is believed that Type~I migration would
dominate. After all, there was a lot more gas than planetesimals in
the original solar nebula. The first work of which we are aware that
addresses this issue was McNeil, Duncan, \& Levison~(2005), which
concentrated on the terrestrial planet region.\footnote{Kominami et
al$.$~(2005) also put Type~I migration and planetesimals in their
studies of terrestrial planet formation. However, they artificially
increased the capture cross-section of their objects by a factor of
25 to decrease accretion time. Unfortunately, this will also
artificially suppress the relative importance of planetesimal-driven
migration.}. They found that, at least for the scenarios they
investigated in this region of the Solar System, planetesimal-driven
migration does not counteract Type~I migration.

However, there are reasons to believe that the situation may be
different in the Jupiter-Saturn zone. For example, the ratio of
$\Sigma_{\rm solid}$ to $\Sigma_{\rm gas}$ is higher because this
region is beyond the snow-line. In addition, the ratio of embryo
surface escape velocity to local circular velocity is larger for an
embryo at $10\,$AU than at $1\,$AU, which might imply that it is a
more effective scatterer. Both of these effects would strengthen the
effects of planetesimal-driven migration.

\begin{figure}
\epsscale{0.8}
\plotone{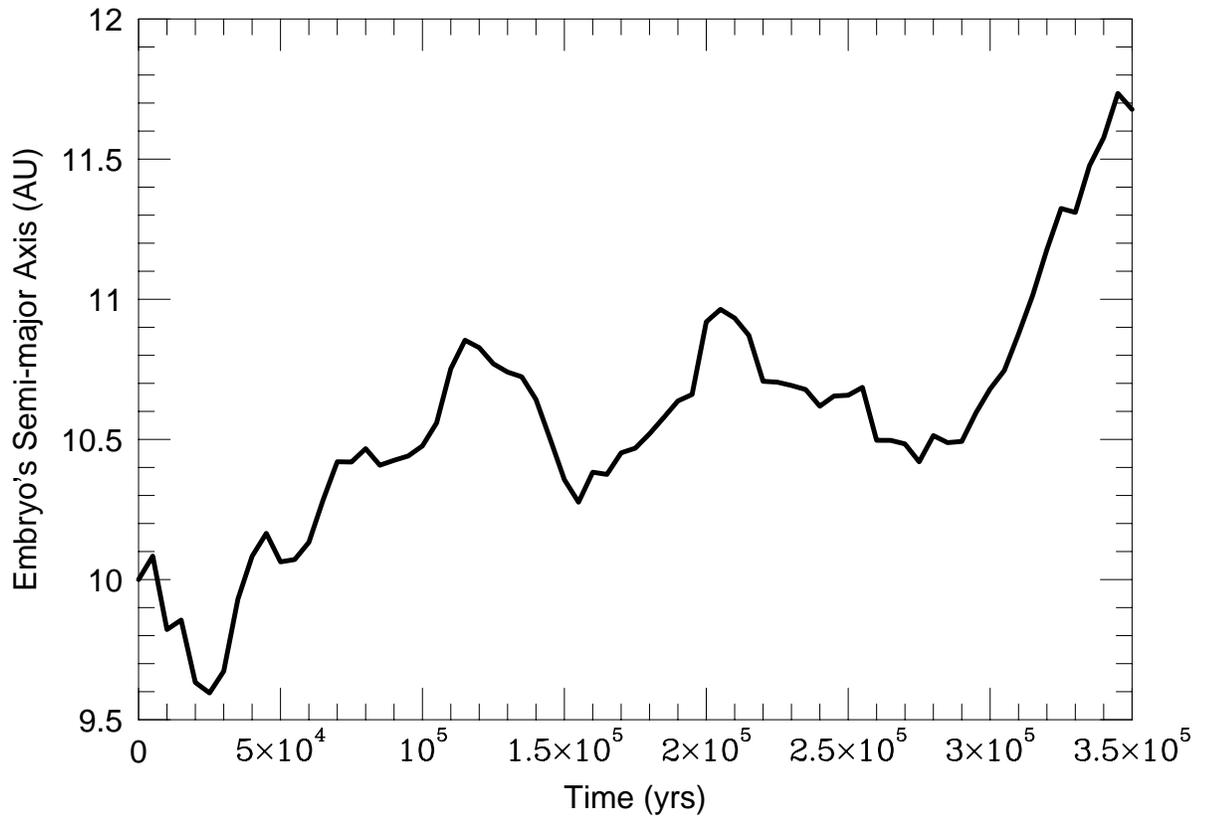}
%\begin{figure}[h!]
%\vglue 3.5truein
%\special{psfile=pmigr.ps angle=0 hoffset=20 voffset=-30 vscale=50 hscale=50}
\caption{\footnotesize \label{fig:pmigr}
The temporal evolution of a Mars-sized object embedded in MMSN disk.
We included the effects of Type~I migration, planetesimal-driven
migration, and aerodynamic drag. Surprisingly, the planet migrates
outward.}
\end{figure}

To test the above hypothesis we ran the following experiment. We
placed a Mars-mass object at $10\,$AU in a MMSN consisting of both a
gaseous disk and planetesimals. The effects of the gas were included
by adding fictitious forces to the equations of motion on the solid
particles. In particular, we added Type~I migration to the planetary
embryo using the formalism in Papaloizou \& Larwood (2000).
\Red{Hydrodynamic} drag was applied to the planetesimals as if they
had a radius of $500\,$m using the methods in Adachi et al$.$~(1976).
In order to save CPU time for this simple experiment, we only placed
planetesimals between 8 and $12\,$AU. The evolution of the embryo's
semi-major axis ($a$) is shown in Figure~\ref{fig:pmigr} --- when
Type~I migration, planetesimal-driven migration, and aerodynamic drag
are included, the planet is transported outward! This behavior is due
to the fact that the gas density is higher closer to the Sun, and thus
aerodynamic drag is stronger and the eccentricity damping timescale is
shorter. So, when the embryo scatters planetesimals, those scattered
inward are more quickly removed from the planet-crossing region than
those scattered outward. In certain circumstances the resultant
asymmetry in angular momentum transferred to the planet can overcome
tendencies for inward migration found in both gas-free planetesimal
scattering and Type~1 migration. As a result, the planet migrates
outward. Thus, at least in principle, planetesimal-driven migration
can overpower Type~I migration in the realm of the giant planets.
Indeed, we investigate this effect more fully in Capobianco et al$.$
(2009).

So, both the results of the above simple experiment and those of gap
opening (McNeil, Duncan, \& Levison~2005; Levison \& Morbidelli 2006)
show that it is important to accurately model the dynamical evolution
of the planetesimal swarm in any calculation of core formation. Only
full $N$-body simulations can accomplish this requirement. Such
simulations are described in the next section.

\section{The Calculations}
\label{sec:code}

Our simulations follow the dynamical evolution of a system containing
a small number of planetary embryos embedded in a sea of
planetesimals. For reasons described \Red{at the end of the last
  section,} these are full $N$-body simulations, where the
trajectories of individual objects are followed as they orbit the Sun
and interact with each other. Our code is based on SyMBA (Duncan et
al. 1998, Levison \& Duncan 2000). SyMBA is a symplectic algorithm
that has the desirable properties of the sophisticated and highly
efficient numerical algorithm known as Wisdom-Holman Map (WHM, Wisdom
\& Holman 1991) and that, in addition, can handle close encounters
(Duncan et al$.$~1998). This technique is based on a variant of the
standard WHM, but it handles close encounters by employing a multiple
time step technique introduced by Skeel \& Biesiadecki~(1994). When
bodies are well separated, the algorithm has the speed of the WHM
method, and whenever two bodies suffer a mutual encounter, the time
step for the relevant bodies is recursively subdivided.

Although SyMBA represented a significant advancement to the
state-of-art of integrating orbits, it suffers from a basic and
serious limitation. At each time step of the integration, it is
necessary to calculate the mutual gravitational forces between all
bodies in the simulation. If there are $N$ bodies, one therefore
requires $N^2$ force calculations per time step, because every object
needs to react to the gravitational force of every other body. Thus,
even with fast clusters of workstations, we are computationally limited
to integrating systems where the total number of bodies of the order
of a few thousand. 

Thus, in order to handle the dynamical evolution of a system
containing a very large number (roughly $10^{26-29}$!) of
planetesimals, some compromises needed to be made. We followed the
techniques in Levison \& Morbidelli~(2006, hereafter LM06), where the
large number of planetesimals is represented by a smaller number of
{\it tracer particles}. Each tracer is intended to represent a large
number of planetesimals on roughly the same orbit as one another. The
embryo-embryo and the embryo-tracer interactions are handled directly.
However, we made some assumptions in order to calculate the
gravitational interaction between the planetesimals. In particular,
close encounters between planetesimals are ignored. This is a
reasonable assumption because the embryos dominate the viscous
stirring of the planetesimals in the \Red{oligarchic regime.  The}
overall gravitational potential of the planetesimals is included using
a 1-dimensional particle-mesh algorithm described in LM06. This
potential must be included in order to correctly handle the resonant
interactions between embryos and planetesimals. Indeed, not including
it leads to unphysical migration of the embryos, which can be
significant when the total mass in planetesimals is comparable or
larger than that of the embryos.

In addition to the gravitational interactions, the code described in
LM06 also models the growth of the embryos via accretion and the
collisional damping of the planetesimals. To accomplish the former,
we assume perfect accretion. Recall that we are following the
dynamical evolution of the embryos and the tracers using direct
$N$-body techniques. Each object is assigned a mass and a radius. If
two objects collide with one another, we simply merge them while
preserving mass, volume, and linear momentum. Note that an important
limitation of this code is that only the embryos can grow; the
planetesimals cannot.

To incorporate collisional damping between planetesimals we adopt the
Monte Carlo techniques described in LM06. In particular, this
algorithm performs a particle-in-a-box calculation on a 1-dimensional
grid of axisymmetric annuli. In order to calculate the collisional
cross-section, the code requires us to specify the radii of the
planetesimals. Following standard practice, we assume that all the
planetesimals have the same radius, $r_{\rm p}$, which is a free
parameter in our models.

The LM06 code was modified to include the additional processes
described in \S{\ref{sec:review}}. We now describe each of these in
detail.

\medskip

\noindent {\it Fragmentation:} The collisional damping algorithm
described \Red{in the last two paragraphs} supplies us with a list of
collisions between the planetesimals during our simulations based on
Monte Carlo techniques.  Included in this information is the impact
velocity of the collision, $v_{\rm imp}$. We use the following Monte
Carlo technique to determine whether a particle fragments.

Recall that each tracer particle actually represents a large number
($N$) of planetesimals of radius $r_{\rm p}$. In particular, $N\!=\!
m_{\rm tr}/ \frac{4}{3}\pi {r_{\rm p}}^3 \rho$, where $m_{\rm tr}$ is
the total mass associated with the tracer, and $\rho$ is the bulk
density, which we assume is $1\,$gm/cm$^{3}$. In order to keep our
calculations tractable, we cannot create new tracers during our
simulation. Thus, for each collision, we are forced to simply assume
that either none of the planetesimals break, or all of them do. If
they break, we assume that the tracer is made up of $N_{\rm f}$
particles of radius $r_{\rm f}$, where $N_{\rm f}\!=\!  m_{\rm tr}/
\frac{4}{3}\pi {r_{\rm f}}^3 \rho$.

Our algorithm is based on the scaling relationships of Benz \&
Asphaug~(1999, hereafter BA99), who determined the fragmentation laws
for objects composed of a variety of different material. In
particular, using hydrodynamic simulations they estimated the ratio of
the mass of the largest fragment resulting from a collision, $m_{\rm
  lr}$, to the total mass of the colliding pair, $m_{\rm tot}$. For
ice (which we assume), they find that
\begin{equation}
\label{eq:mrat}
\frac{m_{\rm lr}}{m_{\rm tot}} \approx -0.6 \left( \frac{v_{\rm
imp}^2}{2Q^{*}_D} - 1\right) + 0.5,
\end{equation}
where $Q^{*}_D$ is the critical specific impact energy needed to
disrupt the target and eject 50\% of its mass. It is also calculated
in BA99. 

Here we use $m_{\rm lr}/m_{\rm tot}$ as a proxy for the probability
that a tracer is fragmented. In particular, for each collision we
calculate this ratio using Eq$.$~\ref{eq:mrat}. Then we generate a
random number between 0 and 1. If the random number is greater than
$m_{\rm lr}/m_{\rm tot}$, we assume that the tracer particle undergoes
fragmentation. Although simplistic, this algorithm has the desired
effect because, on average, it produces the correct ratio of fragments
to planetesimals in our calculations while preserving the total number
of tracers. For example, if we had a collision where $m_{\rm
lr}/m_{\rm tot}=0.9$, we would expect to have 10\% of the mass in
fragments. Our algorithm would disrupt 10\% of such collisions and
thus, on average, 10\% of the mass is also in fragments once we
average over a number of collisions.

So, our code contains three types of particles: 1) the embryos which
are fully interacting and can grow, 2) a population of tracers
representing planetesimals with radius $r_{\rm p}$, and 3) a
population of tracers representing fragments with radius $r_{\rm f}$.
In the current version of the code, the fragments do not collisionally
damp and they cannot fragment again.

\medskip

\noindent
{\it Aerodynamic Drag:} The equations of motion of the planetesimals
and fragments have been modified to include the effects of aerodynamic
drag using the formalism of Adachi et al$.$~(1976). This formalism is
accurate for a wide range of Reynolds and Knudsen numbers, and
includes as limiting cases what are known as the Epstein and Stokes
drag regimes. Our basic algorithm is described in detail in Brasser et
al$.$~(2006). In this case the algorithm has been extended using the
prescription of Adachi et al$.$~(1976) to include cases where the
Knudsen number is larger than unity, which occurs when a molecule's
mean-free-path is larger than the size of the particle. This can occur
for small fragments in the outer region of the nebula.

In order to calculate the drag on particles, we need to adopt
a model for the nebula. Our model, which is based on that of Hayashi
et al$.$~(1985), has the form
\begin{equation}
\label{eq:gden}
\rho_g(\varpi,z) = \rho_{0,g} \left(\frac{\varpi}{1\,{\rm
AU}}\right)^{-\alpha} e^{-z^2/{z_s}^2(\varpi)},
\end{equation}
where $\varpi$ and $z$ are the cylindrical radius and height,
respectively, $\rho_{0,g}$ is the gas density in the plane at $1\,$AU,
and $z_s$ is the scale height of the disk at $\varpi$. The scale
height is determined by the $\varpi$-dependence of temperature $T$:
following Hayashi et al$.$~(1985) we adopt $T=T_0\, (\varpi / {1\,{\rm
AU}})^{-1/2}$ so that
\begin{equation}
z_s(\varpi) = z_{0,s} \left(\frac{\varpi}{1\,{\rm AU}}\right)^{5/4},
\end{equation}
where $z_{0,s}$ is the scale height of the disk at $1\,$AU. Their
``minimum mass '' model has $z_{0,s}=0.047$, $\alpha=2.75$, and
$\rho_{0,g}=1.4\times 10^{-9}\,{\rm gm/cm^3}$. Theoretical arguments
(Lissauer 1987) as well as accretion disk modelling and observational
constraints (reviewed by Raymond et al 2007) suggest that the density
profile was likely shallower and the overall density at 5 AU higher
than suggested by the minimum mass disk model. Thus for the
simulations described here, we adopt $z_{0,s}=0.05$, $\alpha=2.25$,
and $\rho_{0,g}=3.4\times 10^{-9}\,{\rm gm/cm^3}$.

Finally, we need to determine the local circular velocity of the gas,
$v_g$, in our model. As is conventional we define
\begin{equation}
\label{eq:eta}
\eta \equiv \frac{1}{2} \left[ 1 - \left(\frac{v_g}{v_k}\right)^2\right],
\end{equation}
where $v_k$ is the local Kepler velocity. For our assumed temperature profile, 
\begin{equation}
\eta = 6.0 \times 10^{-4} \left(\alpha+\frac{1}{2}\right) \left(\frac{\varpi}{1\,{\rm AU}}\right)^{1/2}.
\end{equation}

\medskip

\noindent {\it Embryo-Disk Tidal Interactions:} For the disk tidal
force exerted on an embryo, we use the approach of Papaloizou and
Larwood (2000), which the authors developed to handle the case where a
protoplanet's eccentricity can be greater than the scale
height-to-semimajor axis ratio. They derive timescales for semimajor
axis damping $t_a$ and for eccentricity damping $t_e$ for an embryo of
mass $M$ at semimajor axis $a$ with eccentricity $e$:

\begin{equation}
\label{eq:ta}
t_a = \frac{1}{c_a} \, 
\sqrt{\frac{a^3}{G M_\Sun}} 
\left(\frac{z_s}{a} \right)^{2} 
\left(\frac{\Sigma_{\rm{g}} \pi a^2}{M_\Sun}\right)^{-1} \,
\left(\frac{M}{M_\Sun}\right)^{-1} \,
\left(
\frac
{1 + (\frac{e a}{1.3 z_s })^{5}}
{1 - (\frac{e a}{1.1 z_s })^{4}}
\right)
\end{equation}

\begin{equation}
\label{eq:te}
t_e = \frac{1}{c_e} \, 
\sqrt{\frac{a^3}{G M_\Sun}} 
\left(\frac{z_s}{a}\right)^{4} 
\left(\frac{\Sigma_{\rm{g}} \pi a^2}{M_\Sun}\right)^{-1} \,
\left(\frac{M}{M_\Sun}\right)^{-1} \,
\left(1 + \frac{1}{4} (\frac{e a}{z_s })^{3}\right)
\end{equation}
where $\Sigma_g$ (= ${\pi}^{1/2} \rho_g z_s$) is the local gas surface
\Red{density.  Papaloizou} and Larwood (2000) also argue that if the
inclination damping timescale \Red{($t_i$)} is not significantly
shorter than the eccentricity damping timescale then it plays little
role in the equilibrium state; we set $t_i = t_e$ for simplicity.

\Red{From} the formulae above we can find the acceleration on an
object due to tidal damping of semimajor axis and random velocity,
namely
\begin{equation}
\label{eq:a}
\vec{a}_{\rm{tidal}} = -\frac{\vec{v}}{t_a} - \frac{2
(\vec{v} \! \cdot \! \vec{r})}{r^2 \, t_e}
- \frac{2 (\vec{v} \! \cdot \! \vec{k})
\vec{k}}{t_i}
\end{equation}
where $\vec{r}$, $\vec{v}$, and
$\vec{a}$ are Cartesian position, velocity, and
acceleration vectors, respectively (with $r$ as the magnitude of the
radial vector) and $\vec{k}$ is the unit vector in the
vertical direction. 

\Red{We adopt} a value of $c_e = 1$, which most researchers agree is
reasonable. There is considerably more controversy about the most
appropriate value for the coefficient $c_a$, since it arises from a
near-cancellation of torques from either side of the planet and may be
strongly affected by nonlinear terms (Paardekooper and Papaloizou
2008). Following the work of many others, we set $c_a = 0$\Red{, which
  implies that we are turning off so called Type~I orbital migration
  in the simulations presented in this paper.}  We will consider the
effects of non-zero type I orbital migration rates in future
publications.

\medskip
\noindent {\it Embryo Atmospheres:} As described in
\S{\ref{sec:review}}, the effective capture cross-section of an embryo
is significantly increased by the presence of an extended atmosphere
that is accreted from the surrounding nebula (II03). In our
calculation, we mimic this effect using the formalism developed by
Chambers~(2006). In particular, assuming the relative velocity of the
particles is small compared to the escape velocity of the embryo
(which is true in our highly damped simulations) and that the scale
height of the atmosphere is set by the energy input due to accreting
planetesimals, the effective accretion radius ($R_{\rm C}$) of an
embyro is
\begin{equation}
\label{eq:rc}
R_{\rm C}^4 = 0.0790\, \frac{\mu^4\, c\,
R^5\, r_H}{\kappa\,
r\, {\dot m}_R} \left(\frac{M}{M_\odot}\right)^2,
\end{equation}
where $R$ and $r$ are the radius of the embryo and planetesimal,
respectively, $M$ is the embryo's mass, $\mu$ is the mean molecular
weight of the atmospheric gas, $\kappa$ is its opacity, and $c$ is the
speed of light. The parameter ${\dot m}_R$ is the accretion rate that
the embryo would have had if there was no atmosphere. We calculate
this value for each embryo in real time during our simulation by
monitoring the number of tracer particles that pass through the
embryo's Hills sphere, and extrapolating to its surface. During our
simulations, we did not allow $R_{\rm C}$ to exceed $0.5\,r_H$.

\begin{figure}
\epsscale{1.}
\plotone{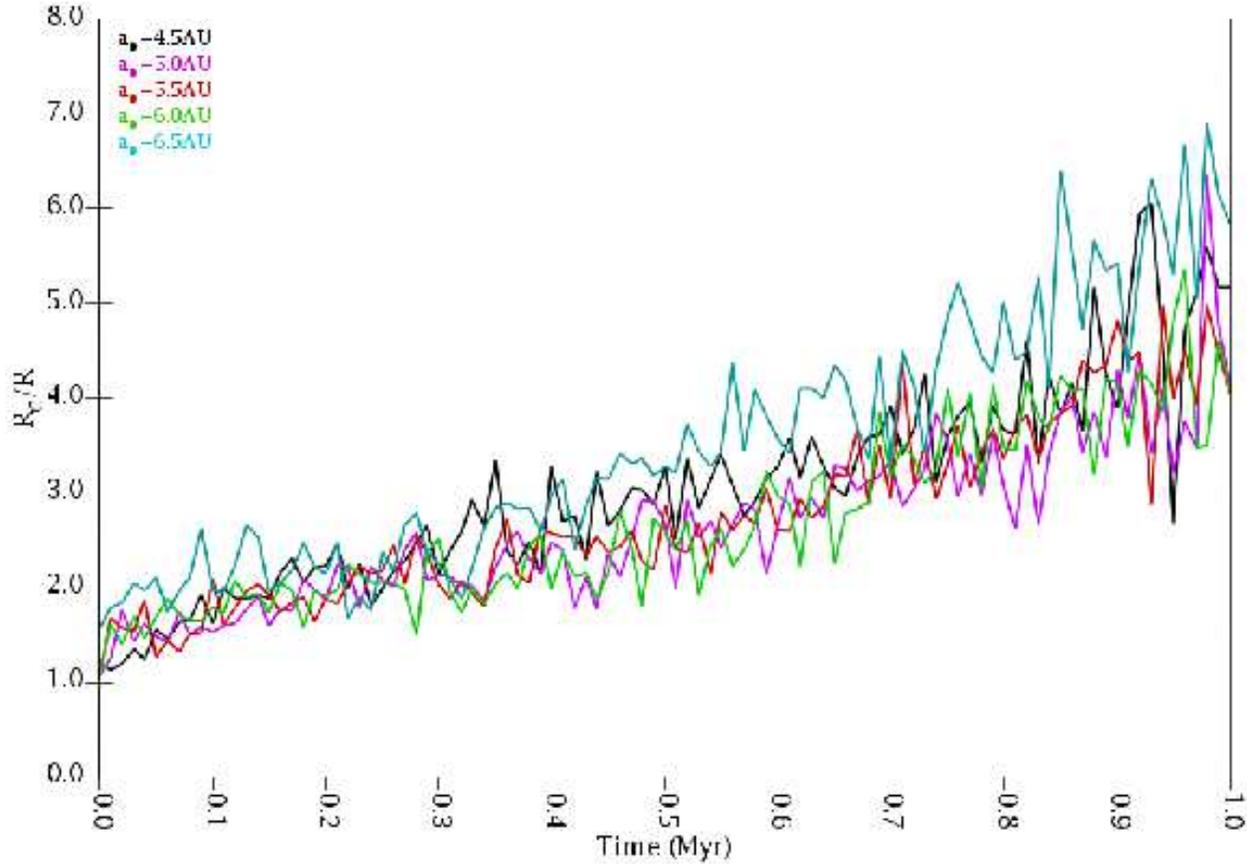}
%\begin{figure}[h!]
%\vglue 4.0truein
%\special{psfile=mdot.ps angle=0 hoffset=10 voffset=0 vscale=70 hscale=70}
\caption{\footnotesize \label{fig:rc}
The ratio of $R_C$ to $R$ for the 5 embryos in the first simulation
discussed in \S{\ref{sec:results}}. Each embryo is represent by a
different color and is identified by its initial semi-major axis,
$a_0$ as shown in the legend.}
\end{figure}

As first described in II03, the existence of an extended atmosphere
can significantly enhance the capture cross-section of the embryos.
For example, Fig$.$~\ref{fig:rc} shows $R_C/R$ for the five embryos in
the first simulation presented in the next section during the initial
million years of evolution. These embryos are initially $1\,M_\oplus$
and grow to an average of $2.1\,M_\oplus$ during the time period
shown. The increase in $R_C/R$, however, is due to a decrease in the
accretion rate, which occurs because the embryos clear their feeding
zone (see discussion below). As can be seen from the figure, $R_C$ can
be almost an order of magnitude larger than $R$ in this simulation and
the enhancement factor is even larger is some of the other runs.

\medskip
\medskip

We next discuss the initial conditions for the majority of our
simulations. As described in \S{\ref{sec:intro}}, our goal is not to
perform complete simulations of giant planet core formation, but to
evaluate the effectiveness of various dynamical mechanisms that might
aid in the formation process. As a result, our simulations are
somewhat idealized.

We follow the evolution of a system of five Earth-mass embryos
embedded in a disk of planetesimals with the code described above.
The embryos occupy the region between 4.5 and $6.5\,$AU so that they
are separated by roughly 10 Hill radii. They initially have
eccentricities of 0.002 and inclinations of $0.05^\circ$. The
orientation of their orbits were chosen at random.

These embryos were places in a planetesimal disk that stretched from 4
to $16\,$AU. Note that this implies that the embryos were at the inner
edge of the disk. We designed our simulations in this way so that the
disk is capable of supplying additional planetesimals to the region of
embryo growth if physical conditions allow. The disk initially was
$200\,M_\oplus$, which is six times the minimum mass solar nebula in
this region (Hayashi et al$.$~1985). Previous models (Lissauer 1985;
TDL03) have argued that this amount of material is needed to grow the
cores. The disk was represented by 20,000 tracer particles.  Their
eccentricities and inclinations were generated from Raleigh
distributions with $e_{\rm RMS}=0.01$ and $\sin{(i)}_{\rm RMS}=0.005$.
In the next section we describe the dynamical and collisional
evolution of this system that occurs as a result of the processes
described above. Unless otherwise noted, the simulations were run for
$3\,$Myr, in accordance with estimates for the lifetimes of the
gaseous disks (Haisch et al$.$~2001). However, we held the density of
the gas constant during these calculations.

\section{Simulation Results}
\label{sec:results}

As described in the previous section, our simulations follow the
growth of five Earth-mass planetary embryos in the region near
Jupiter's current orbit. Our goal is to determine which combination of
effects and parameters will allow these Earth-mass embryos to grow to
$\sim\!10\,M_\oplus$ so that they can accrete a massive gaseous
atmosphere similar to those of Jupiter and Saturn. Our simulations
include a wide range of physical parameters that have yet to be
included in any full $N$-body simulation. The five parameters that we
varied are described in detail in \S{\ref{sec:code}} and summarized in
Table~\ref{tab:free}. A five dimensional parameter space is too large
to cover uniformly. Thus, we probed this space strategically ---
investigating directions in parameter space that seemed likely to
elucidate the roles of the physical processes we are interested in. In
all we performed 172 simulations --- only a subset of the most
illustrative will be discussed below.

\begin{table}[h]
\singlespace
\begin{center}
\begin{tabular}{|ll|}
\hline
$r_{\rm p}$: & Radius of planetesimals \\
\hline
$r_{\rm f}$: & Radius of fragments \\
\hline
$\rho_{0,g}$: & The mid-plane gas density of the nebula at $1\,$AU\\
\hline
$c_e$: & Type I eccentricity decay coefficient\\
\hline
$c_a$: & Type I orbital decay coefficient\\
\hline
$\kappa$: & Opacity of nebular gas\\
\hline
\end{tabular}
\caption{\label{tab:free} A list of the free parameters in our simulations.}
\end{center}
\end{table}

We first describe a series of runs intended to investigate the effects
of aerodynamic drag on the growth of the embryos. In particular, we
studied systems where we turned off fragmentation and varied $r_{\rm
  p}$ from 1 to $100\,$km with a spacing of 0.5 dex. We set
$\rho_{0,g}\!=\!3.4\times 10^{-9}\,$g/cm$^3$, which produces a solar
solid-to-gas ratio for the overall disk. For reasons described above,
we set $c_a\!=\!0$, but allowed the gas to damp the eccentricities of
the embryos by setting $c_e\!=\!1$. And finally, we set $\kappa$ to
2\% of the standard interstellar medium value, in accordance with
Hubickyj et al$.$~\Red{(2005)}. We performed 15 simulations for each
value of $r_{\rm p}$. Each of these had slightly different initial
conditions.  In addition, we changed the seed for the random number
generator.

\begin{figure}
\epsscale{1.}
\plotone{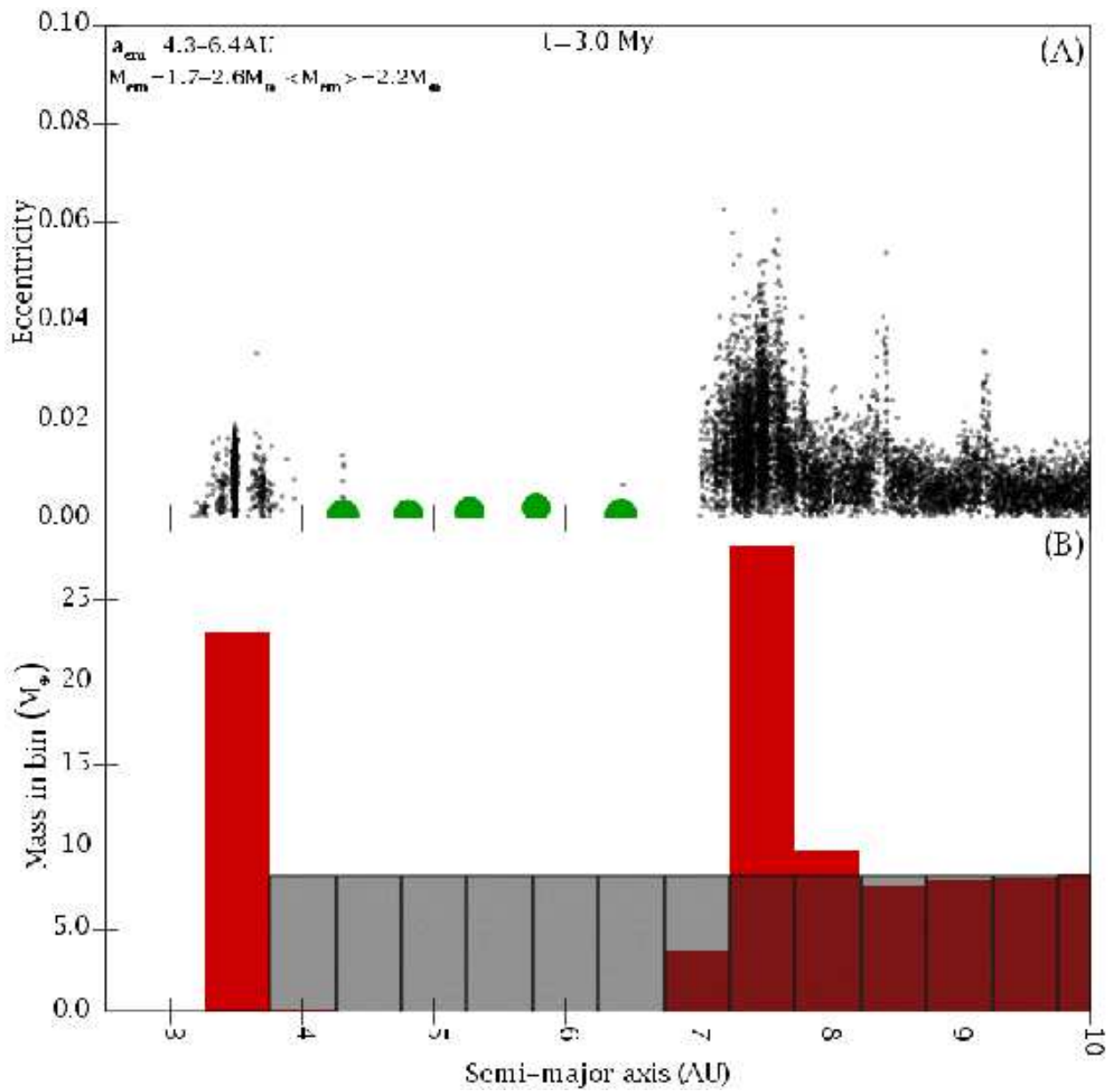}
%\begin{figure}[h!]
%\vglue 5.1truein
%\special{psfile=aesig_2-100km-no.ps angle=0 hoffset=0 voffset=0 vscale=70 hscale=70}
\caption{\footnotesize \label{fig:run0}
The state of the system in Run~A at $3\,$Myr. A) Eccentricity as a
function of semi-major axis. The green circles show the embryos,
while the black dots are the planetesimals. The size of the green
circles are proportional to the cube root of the embryo mass. There
are no fragments in this figure (which are normally red) because
fragmentation is disabled in this simulation. (B) The distribution
of planetesimal mass. In particular, this shows the amount of mass
in annular bins that have a width of 0.5 AU. The gray histogram
shows the distribution at the beginning of the simulation, while the
red shows it at the end.}
\end{figure}

We found that 90\% of our simulations exhibited similar behavior.
Figure~\ref{fig:run0} shows the state of such a system ($r_{\rm
p}\!=\!100\,$km, hereafter Run~A) at $3\,$Myr. Recall that our goal
is to grow the embryos to $\sim\!10\,M_\oplus$. In this simulation
the embryos range in mass from 1.7 to $2.6\,M_\oplus$. Note that the
region occupied by the embryos has been cleared of planetesimals
(except at the embryos' Lagrange points) and thus we should not expect
any more growth. Indeed, all the growth in this simulation occurred in
the first million years.

The fact that both the region surrounding the embryos is empty and the
embryos only accreted a total of $3.3\,M_\oplus$ might seem surprising
given that this region originally contained $49\,M_\oplus$ of
planetesimals. The reason for this can be seen in
Figure~\ref{fig:run0}B, which shows the distribution of planetesimals
at $t\!=\!0$ (gray histogram) and at $t\!=\!3\,$Myr (red histogram).
The planetesimals originally between the embryos were scattered away
rather than accreted. Aerodynamic drag then removed them from the
embryos and placed them on \Red{orbits that protects them from
  encountering the cores.}  This result shows the importance of
including real $N$-body effects in any study of giant planet core
formation. Previous modelling attempts (Hubickyj et al$.$~2005;
Alibert et al$.$~2005; Chambers~2008), which use semi-analytic
approaches to model the embryo-planetesimal interactions, do not allow
the embryos to redistribute the planetesimals. Clearly this process is
important even when the cross-sections of the embryos are enhanced by
an atmosphere.

The dynamical evolution described above did occasionally lead to the
construction of $>\!10\,M_\oplus$ cores. \Red{In} our simulations we
always develop dense, dynamically cold rings of planetesimals
immediately exterior to the region containing the embryos (c$.$f$.$
Fig$.$~\ref{fig:run0}B). In a few of our simulations the embryos
jostle each other enough so the one of the embryos is pushed into one
of the rings where it can grow. When this happens the embryo quickly
grows $\sim\!10\,M_\oplus$. This primarily occurs in simulations with
$r_{\rm p}\!=\!1\,$km.

Unfortunately, we believe that the \Red{growth mechanism in the last
  paragraph} probably would not occur in nature, at least to the
extent that we observe \Red{in our calculations}, and is a result of
the simplifications that we employ. In the above integrations, rapid
growth occurs when an embryo is scattered into a dense, dynamically
cold, ring of planetesimals.  Such rings form during calculations
where damping is large (i$.$e$.$ $r_{\rm p}$ is small) in regions that
lack the embryos. These rings remain dynamically cold because our
algorithm does not allow the planetesimals to dynamically excite one
another. In addition, we do not let the planetesimals accrete with one
another and thus build another embryo, which would further excite the
rings. \Red{Since we do not see this rapid growth} in runs where the
rings do not suffer from a large amount of damping, we think that a
more realistic treatment of planetesimal-planetesimal interactions
would effectively squelch this mode of growth.

\subsection{The Role of Planetesimal Driven Migration}
\label{ssec:aero}

While 90\% of the runs in this series followed the behavior where the
embryos clear their immediate region and very little growth occurs,
the remaining 10\% show a new, important mechanism in core formation.
This process regularly built $>\!10\,M_\oplus$ embryos and thus is our
first success at producing reasonable giant planet cores. In these
simulations the embryos undergo a self-sustaining outward migration
driven by the gravitational interactions with the planetesimals. This
leads to significant growth. An example, which we call Run~B, is
presented in Fig$.$~\ref{fig:rawayf}. This example has the same
parameters as Run~A, except that $r_{\rm p}\!=\!10\,$km. The embryo
that ends up the farthest from the Sun grows from $2\,M_\oplus$ to
$20\,M_\oplus$ in only $\sim\!10^5$ years. Note that our simulations
do not include the direct accretion of nebular gas, and so this
$20\,M_\oplus$ embryo is composed entirely of solids. During the same
period, it migrates from $7$ to $12\,$AU.

\begin{figure}
\epsscale{1.}
\plotone{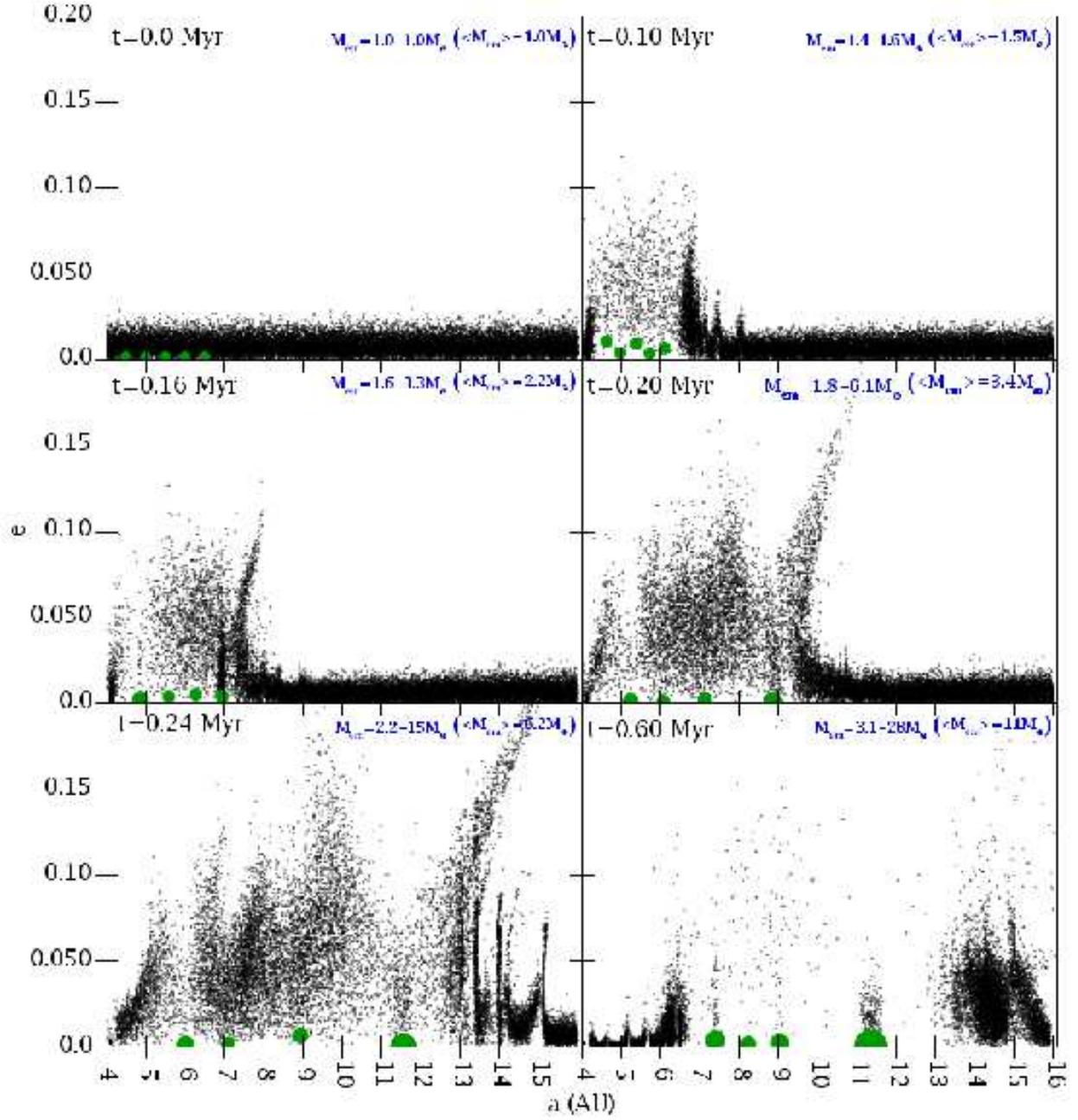}
%\begin{figure}[h!]
%\vglue 7.0truein
%\special{psfile=PS/ae_run.ps angle=0 hoffset=-50 voffset=-90 vscale=90 hscale=90}
\caption{\footnotesize \label{fig:rawayf}
  The temporal evolution of Run~B, which suffers self-sustaining
  outward migration. See the caption for Fig$.$~\ref{fig:run0}A for an
  explanation of these panels.}
\end{figure}

The system evolves in the following manner. The five embryos are
initially in nearly circular orbits and separated by $10\, r_H$. They
start to grow by eating the nearby planetesimals. Despite the
inclusion of dynamical friction with the planetesimals and
eccentricity damping due to the gaseous disk, this growth allows the
embryos to gravitationally excite one another until they begin to
suffer close encounters. At 140,000 years, the inner two embryos merge
with one another. In addition at roughly the same time, the second
furthest embryo is scattered outward (this appears to be independent
of the merger). It then becomes the most distant embryo, but more
importantly it penetrated a high density region of the disk.  This
triggers a period of fast migration and growth. At the end of the
simulation \Red{(i.e$.$ at $3\,$Myr)}, in order of distance from the
Sun, the embryos are 6.9, 3.1, 5.3, and $28\,M_\oplus$.

We performed 15 simulations for each value of $r_{\rm p}$. In all the
runs that underwent self-sustaining migration, the outer embryo was
always the largest \Red{and the remaining embryos did not} grow to
$10\,M_\oplus$. There is a rough relationship between the mass of the
largest embryo in a simulation and $r_{\rm p}$. For $r_{\rm p}\!=\!3$,
10, 30, and $100\,$km, the average largest embryos were 27, 24, 13 and
$13\,M_\oplus$, respectively.  This is probably due to the fact that
there is more damping in systems with smaller $r_{\rm p}$ and thus
accretion is more efficient.

Although not statistically significant, there is also a noticeable
trend between the fraction of runs that underwent self-sustaining
outward migration and $r_{\rm p}$. In particular, four of the
simulations with $r_{\rm p}\!=\!10\,$km experienced outward migration,
and the number drops off for either larger or smaller radii. This may
indicate that outward planetesimal migration could be aided by
aerodynamic drag.

In brief, as \Red{a planet scatters} planetesimals, there are loss
mechanisms or sinks both interior to or exterior to its orbit (see the
review by Levison et al$.$~2007). These sinks can dominate the
embryo's migration and overcome its tendency to migrate inward due to
a local asymmetry in outward versus inward scattering (Kirsh et
al$.$~2009).  If the interior (exterior) sinks are more important then
the embryos will move outward (inward) to conserve angular momentum.
At the beginning of migration when an embryo is deciding which
direction to go, the only sinks \Red{for its planetesimals (in the
  absence of gas) are passing them off to its brethren, or ejecting
  them} from the system.  However, when gas is present, it can be the
main factor in determining the direction of migration.  Because the
density of the gas is a strongly decreasing function of heliocentric
distance, an embryo is more likely to lose a planetesimal if it
scatters the small object inward where the gas density is highest.
That is, the embryo scatters the planetesimals inward, where the gas
circularizes the orbit so that it is beyond the reach of the embryo.
As a result, there is a major sink interior to the orbit of the embryo
that, under the \Red{appropriate} conditions (combination of
$\rho_g(\varpi,z)$ [Eq$.$~\ref{eq:gden}] and $r_{\rm p}$) can trigger
outward migration.  Important insight into this process can be gained
by studying the one-planet case. We do this in Capobianco et
al$.$~(2009).

Given the above result, it is interesting to inquire about how much of
the growth seen in the outer embryo is the result of its extended
atmosphere. Thus, we performed a series of 15 simulations with the
same initial conditions as those above (c.f$.$
Fig$.$~\ref{fig:rawayf}), but without the atmosphere. We found
self-sustaining outward migration in only one of these simulations.
While 1 out of 15 is not necessarily statistically distinct from the 4
out of 15 found above, the character of the migration is different.
Like the run shown in Fig$.$~\ref{fig:rawayf}, there was a burst of
migration that pushed the outermost embryo from 7 to beyond $11\,$AU
in only $60,000$ years. The major difference between this and previous
runs is that during the migration the embryo grew by only
$0.6\,M_\oplus$. Indeed, by the end of the simulation the largest
embryo was only $3.5\,M_\oplus$. Thus, we tentatively conclude that a
massive gaseous envelope greatly increases the probability of
substantial growth of the embryos during planetesimal driven
migration.

Given the success of our simulations that combine massive atmospheres
and planetesimal driven migration, we must address the fact that only
$\sim\!10\%$ of our simulations underwent outward migration before we
can conclude that this mechanism is a viable solution to the core
formation problem. We believe, however, that this process might be
more robust than the above models suggest. For reasons described
previously, the above simulations contain many simplifications.
Perhaps most important for this issue is that both the embryos and the
planetesimals are represented by objects of a single size. We can
imagine that, for example, if the planetesimals were represented by a
more realistic size-distribution, there might routinely be enough
objects at the correct size to cause this outward migration.
Unfortunately, our code cannot handle this situation (see however,
Capobianco et al$.$ 2009), but it can currently accommodate embryos of
different initial masses. Thus, we performed a series of simulations
with the same parameters as the Run~A and with $r_{\rm p}$ was either
1, 10, or $100\,$km. In addition to the 5 Earth-mass embryos, however,
we added a population of 10 additional $0.1\,M_\oplus$ (i$.$e$.$ {\it
  $\sim$Mars-mass}) embryos. The initial conditions for these
simulations are shown in Fig$.$~\ref{fig:mars}A.

\begin{figure}
\epsscale{1.}
\plotone{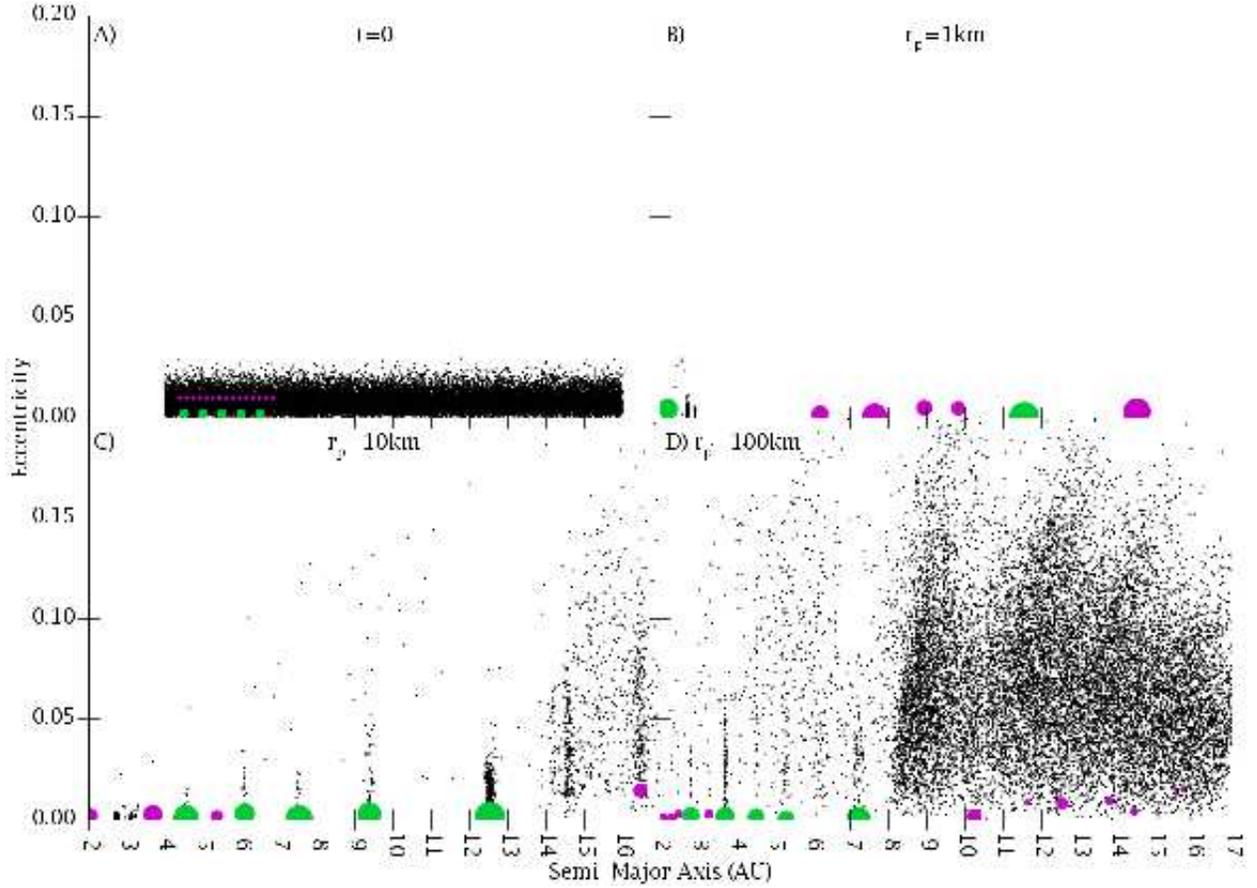}
%\begin{figure}[h!]
%\vglue 4.5truein
%\special{psfile=PS/ae_mars.ps angle=0 hoffset=-10 voffset=-150 vscale=80 hscale=80}
\caption{\footnotesize \label{fig:mars}
A) The initial conditions for the simulations that contain
$\sim$Mars-mass embryos. See Figure~\ref{fig:run0} for a
description of the colors and symbols. Here, the green represents
the initial Earth-mass embryos, while the purple shows the initial
$\sim$Mars-mass embryos. B) $r_{\rm p} = 1\,$km at $3\,$Myr. C)
$r_{\rm p} = 10\,$km at $3\,$Myr. D) $r_{\rm p} = 100\,$km at
$3\,$Myr.}
\end{figure}

Before we describe the results of these calculations, we first must
issue a warning. As described above, in these calculations we use a
population of 20,000 tracer particles to act in place of a much larger
number of planetesimals. Previous studies (e.g$.$ Levison \&
Morbidelli~2007) have shown that this representation is valid as long
as the embryo to tracer mass ratio is $\gtrsim 100$. This is true for
our main simulations where the mass ratio is precisely this value.
Indeed, it is for this reason that we restricted our main studies to
Earth-mass embryos in the first place. However, in the experiments we
are about to discuss, the ratio is only 10. Thus, the $\sim$Mars-mass
embryos are experiencing significantly larger Brownian motion than a
more realistic simulation would produce.  (Unfortunately, it is not
practically feasible to significantly increase the number of tracers.)
Thus, the results of these simulations should be viewed with caution.
Have said this, we believe that these simulations show that
self-sustaining outward migration might be significantly more robust
than the simple simulations discussed above would imply, and thus we
briefly discuss them.

Figs$.$~\ref{fig:mars}B--D show each system at $3\,$Myr. In both the
$r_{\rm p}\!=\!1$ and $10\,$km calculations self-sustaining outward
migration is clearly apparent because there are objects that were
originally Earth-mass embryos (green in the figure) at 11.6 and
$12.5\,$AU, respectively. Outward migration also occurred in the
$r_{\rm p}\!=\!100\,$km simulation, but the embryos migrated back in.
Indeed, the embryo started at $6.5\,$AU, migrated to $10.5\,$AU in
$10^5\,$yr, migrated back to $6.1\,$AU, out to $8.5\,$AU, back to
$6.0\,$AU, and at the end of the simulation is migrating outward
again. Therefore, rather then being a rare occurrence, outward
migration is the norm in these simulations.

The big embryos migrate because of the effects of their smaller
relatives on the planetesimals. In all cases we see that, although
the $\sim$Mars-mass embryos were initially distributed among their
larger brethren, a significant fraction are scattered out of this
region on the timescale of a few $\times 10^4$ years. Most of these
evolve onto nearly-circular orbits just outside of it. Recall that in
Run~A (Fig$.$~\ref{fig:run0}) many of the planetesimals scattered from
between the embryos are also trapped in this region. It is not
surprising that the $\sim$Mars-mass embryos behave like the
planetesimals when they interact with the Earth-mass embryos because
in both cases the mass ratio is large.

\Red{It is important to note that} in these runs, the $\sim$Mars-mass
embryos that come to populate the region beyond the Earth-mass objects
actually feed planetesimals back inward. This process causes some of
Earth-mass embryos to migrate outward. The smaller embryos are pushed
outward as well. Thus, the whole system spreads and the embryos grow,
resulting in the systems seen in Figs$.$~\ref{fig:mars}B--D. In each
of these three simulations, objects originally $1\,M_\oplus$ grew to
$>10\,M_\oplus$ --- the largest being $28\,M_\oplus$. Indeed, four of
the five big embryos in the $r_{\rm p}\!=\!10\,$km run grew larger
than $>10\,M_\oplus$, and the fifth was $8.7\,M_\oplus$.

\Red{The issue now} is how to interpret the success of these
simulations given the above warning. The main difference between these
runs and the ones with five embryos is that the large cores deliver
the $\sim$Mars-mass objects to a region where they can feed
planetesimals back inward. We see no reason why the low-resolution of
the disk should effect this evolution. Thus, we feel that this part of
the simulations would indeed happen in a more realistic model.

\Red{We} believe that these calculations show that self-sustaining
outward migration would commonly result if we could include a spectrum
of embryo sizes. Indeed, systems with a strong size gradient are a
probable outcome of the earlier phases of planet formation: analytic
models of oligarchic growth (Thommes et al$.$ 2003) predict that prior
to \Red{growing} to their isolation masses, the embryo masses should
vary as $a^{-X}$, with $X \sim 6$ for the surface density distribution
we use here. Thus one expects that at the time that an oligarchic
embryo approaches a mass of $ \sim 1\,M_\oplus$ near 5 AU, embryos
beyond 7.5 AU will each be more than an order of magnitude smaller in
mass. \Red{Again, we} use similar size embryos between 4.5 and 6.5 AU
in our main simulations \Red{in order to adequately resolve the the
  embryo-disk interactions.}  It still remains to be demonstrated
whether a more realistic size spectrum will actually lead to
$10\,M_\oplus$ cores.  Unfortunately, we must leave this work to the
future.

\subsection{The Role of Fragmentation}
\label{ssec:frag}

As described in \S{\ref{sec:review}}, it is believed that
fragmentation will aid core formation because the small remnants of a
\Red{planetesimal-planetesimal} collision will damp quickly and thus
be more likely to be accreted by the growing embryos (Wetherill and
Stewart 1993; R04). One of the main motivations of this work is to
test this hypothesis.

\begin{figure}
\epsscale{1.}
\plotone{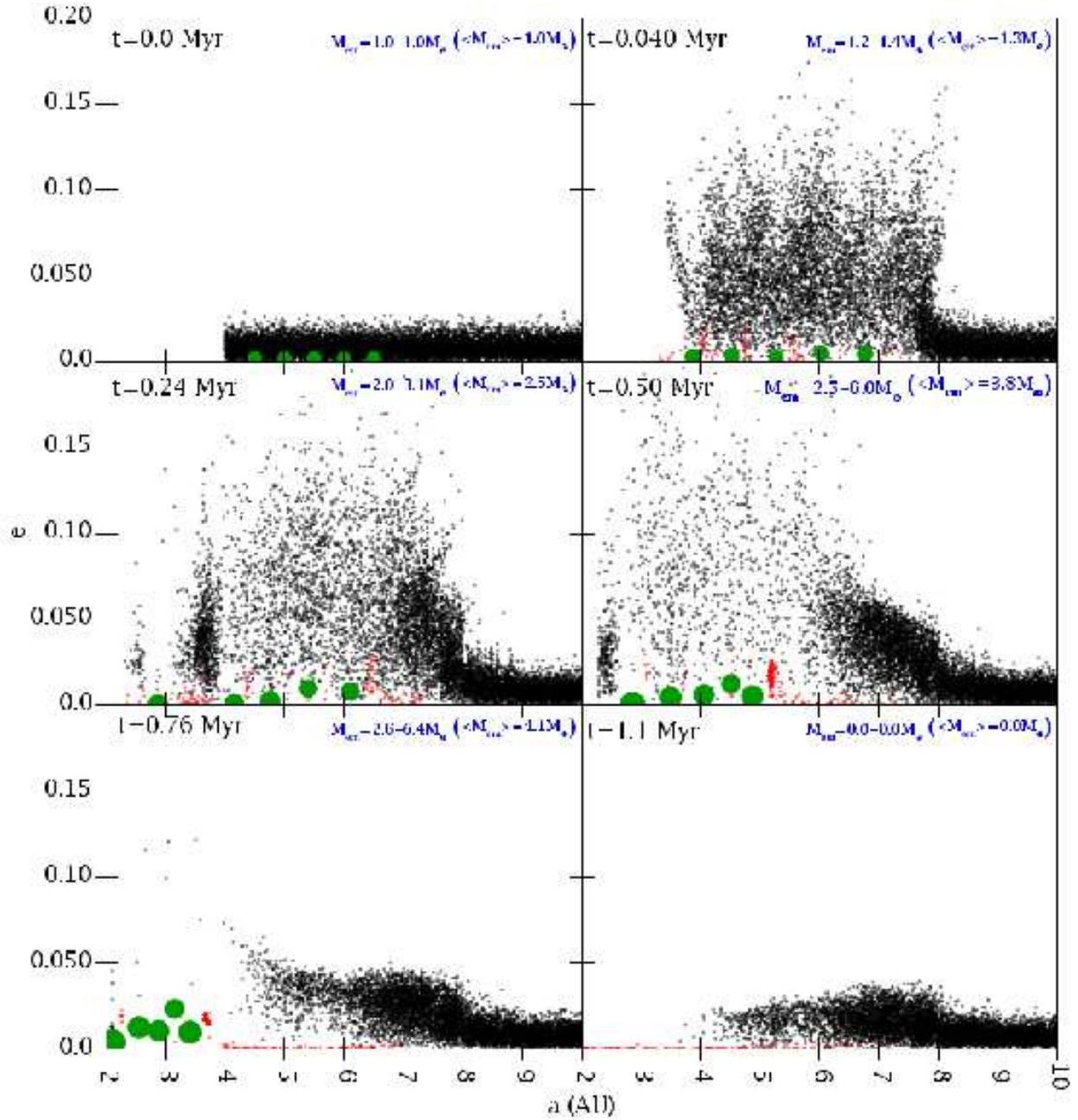}
%\begin{figure}[h!]
%\vglue 7.0truein
%\special{psfile=PS/ae_push.ps angle=0 hoffset=-50 voffset=-90 vscale=90 hscale=90}
\caption{\footnotesize \label{fig:push}
  The temporal evolution of a system like Run~A, but in which
  fragmentation is turned on. See the caption for
  Fig$.$~\ref{fig:run0}A for an explanation of these panels. Note that
  the red dots represent fragments of collisions.}
\end{figure}

Fig$.$~\ref{fig:push} shows the evolution of a system with the same
characteristics as Run~A under the influence of fragmentation with
$r_{\rm f}\!=\!100\,$m. Early on, the system behaves as one would
expect. The embryos dynamically excite the planetesimals (black dots
in the figure). This leads to collisions which fragment the
planetesimals. Due to their much smaller size, the fragments (red
dots) have much smaller eccentricities than the planetesimals (the RMS
eccentricity is 0.006 as opposed to 0.03 at $t\!=\!40,000\,$yr).
Also, they spiral inward at a much faster rate. For the aerodynamic
drag models described in \S3, Adachi et al$.$ (1976) show that gas
drag will cause a fragment with physical density $\rho_f$,
eccentricity $e$ and inclination $i$ to decay in semimajor axis $a$ at
a rate given for small $e$ and $i$ by
\begin{equation}
\label{eq:dadt}
\frac{da}{dt} = - \frac{2 a \eta}{\tau_0}\left(0.97 e + 0.64i + \eta\right),
\end{equation}
where \Red{$\eta$, which is defined in Eq$.$~\ref{eq:eta}, is a
  function of the ratio between the circular velocity of the gas and
  the local Kepler velocity, and}
\begin{equation}
\tau_0= \frac{8 \rho_f r_f}{3 C_D \rho_g v_k}\,\, .
\end{equation}

Thus, near 5 AU in the nebular model we have adopted, 100 m fragments
on near-circular orbits ($ e,i \ll \eta$ ) would be expected to
migrate inward at roughly $2 \times 10^{-5}$ AU/yr. However, rather
than sweeping by or being accreted by the embryos, the fragments
typically become trapped in mean motion resonances. For example, the
clump of fragments near $5.2\,$AU in the $t\!=\!500,000\,$yr panel is
in the 11:10 mean motion resonance with outer embryo. This clump
contains a total mass of $1.3\,M_\oplus$.

Once in resonances, the embryos try to stop the inward migration of
the fragments (Weidenschilling \& Davis~1985). However, the gas
continues to draw angular momentum from the fragments' orbits.
\Red{The} interaction between the resonance and the aerodynamic drag
pumps up the eccentricities of the resonant particles in the clump (to
values of $\sim 0.015$, consistent with the prediction of
$[\eta/(j+1)]^{1/2}$ for a $j/(j+1)$ resonance by Weidenschilling and
Davis 1985). The larger eccentricity increases the rate at which the
fragments attempt to spiral in to $\sim 10^{-4}$ AU/yr (cf$.$
Eq$.$~\ref{eq:dadt}). However, the resonance lock means that the
angular momentum loss is shared with the interior planet, and by
virtue of that body's interactions with the planets interior to it,
with the whole retinue of planets. Thus, although the drag acts
directly on the $1.3\,M_\oplus$ of fragments, the whole ensemble of
$\sim 22 \,M_\oplus$ "shares" the angular momentum loss and the group
migrates in at roughly $1/20$ of the rate predicted for fragments of
that eccentricity. Consequently, by 1 million years, the embryos are
pushed inside of $2\,$AU.\footnote{The timestep of our calculations
  \Red{is} determined by the shortest orbital period in the problem.
  Thus, in order to keep the timestep large enough to make the
  calculations practical, we removed any object \Red{that} got closer
  to the Sun than $2\,$AU.}

A note of caution is now in order. As we described above, we do not
allow fragments to collisionally interact with one another in these
calculations. The embryo migration rate is determined, in part, by the
eccentricities of the fragments. This, in turn, \Red{is} dependant on
the eccentricity damping rate, which is, in reality, affected by both
aerodynamic drag and collisions. \Red{By} not including fragment
collisional damping, we are overestimating the eccentricity and thus
the embryo migration rate. In addition, there is the potential that a
collision between two resonant objects could knock \Red{them} out of
the resonance. In order to estimate these effects, we have performed
simulations of a system containing one embryo and a series of
particles in the embryo's 7:6 mean motion resonance (which is commonly
populated in our main simulations). We find that collisions were very
inefficient at removing objects from the resonance. Indeed, the
libration amplitudes of the particles decreased during the simulation.
\Red{The} migration rate of the embryo in the calculation without
collisions is only 1.6 times larger than that with collisions.  Thus,
although the effect of collisions is significant in the quantitative
sense, \Red{we believe} that the fact that it was not included
\Red{does not invalidate} our basic conclusion that resonant trapping
can lead to wholesale inward migration of the growing cores.

Earlier work (Weidenschilling \& Davis~1985; Kary et al$.$~1993) also
tells us that not all small objects undergoing radial drift due to
aerodynamic drag will be caught in resonances. The probability of
capture is determined, in part, by the strength of the drag
acceleration. So, our conclusion above might not be robust in that
there might be large regions of parameter space in which resonant
trapping does not occur. To test this, we performed a series of
simulations where we varied $r_{\rm f}$ from 1 to 100 meters with a
resolution of $\sim\!0.5$ dex. We performed integrations with $r_{\rm
  p}$ set to both 10 and $100\,$km --- the former chosen in the hope
that planetesimal driven outward migration could negate the effects of
resonant trapping. We found that for $r_{\rm f}\!\geq\!30\,$m all the
embryos were lost by being pushed inward by the fragments.

% We
%found that for $r_{\rm f}\!>\!30\,$m and for the $r_{\rm
% f}\!=\!30\,$m$, r_{\rm p}\!=\!10\,$km run all the embryos were lost
%by being pushed inward by the fragments.

\begin{figure}
\epsscale{1.}
\plotone{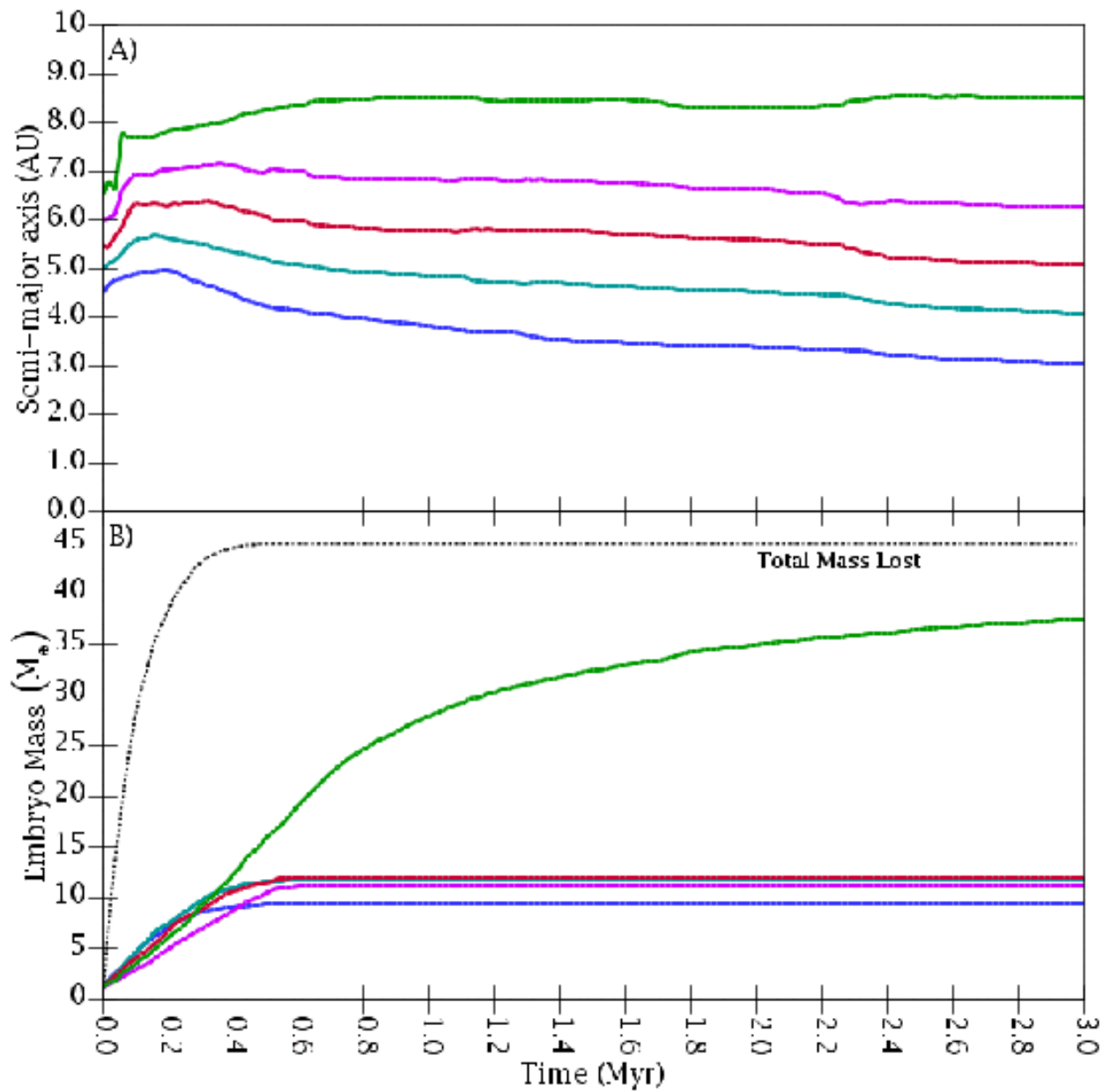}
%\begin{figure}[h!]
%\vglue 6.6truein
%\special{psfile=PS/aM_raf.ps angle=0 hoffset=-40 voffset=-90 vscale=85 hscale=85}
\caption{\footnotesize \label{fig:raf} The temporal evolution of the
five embryos in the run with $r_{\rm f}\!=\!10\,$m, $r_{\rm
p}\!=\!10\,$km, and $\kappa$ to 2\% of the standard interstellar
medium value. A) The semi-major axes as a function of time. Each
embryo is represented by a solid curve of a different color. B) The
mass as a function of time. The solid curves show the embryos,
where the color corresponds to the semi-major axes in (A). The
dotted curve show the mass lost from the system because it got too
close to the Sun.}
\end{figure}

The runs with $r_{\rm f}\!\leq\!10\,$m behaved as the analytic and
numerical experiments models cited \Red{in the last paragraph} found.
That is, fragments were created in planetesimals collisions, were
damped by the gas, spiraled inward, and were accreted by the embryos
while in very low-inclinations orbits (i.e$.$ in the shear regime) or
migrated right past, as happened most often for the smallest
fragments. No resonant trapping was observed because the fragments
were migrating faster than the critical value at which trapping begins
to be ineffective (Kary et al$.$ 1993).

\Red{We} find that this process only allows for a significant amount
of growth for a narrow range of $r_{\rm f}$ --- in particular when
$r_{\rm f}\!=\!10\,$m for the gas disk adopted here. In order to
understand this, we must first discuss the case where the embryos grew
substantially. Fig$.$~\ref{fig:raf} shows a simulation where all 5
embryos reach a mass of nearly $10\,M_\oplus$ (the inner one is
actually $\sim\!9.5\,M_\oplus$, but the rest were $>\!10\,M_\oplus$).
In this run, which we call Run~C, all the embryos enjoy a roughly
linear growth in mass for the first $\sim\!400,000$ years. After this
time, growth of all but the outermost embryo stops. The outer core
continues to grow \Red{until it is $38\,M_\oplus$ at the end of the
  simulation.}

\begin{figure}
\epsscale{1.}
\plotone{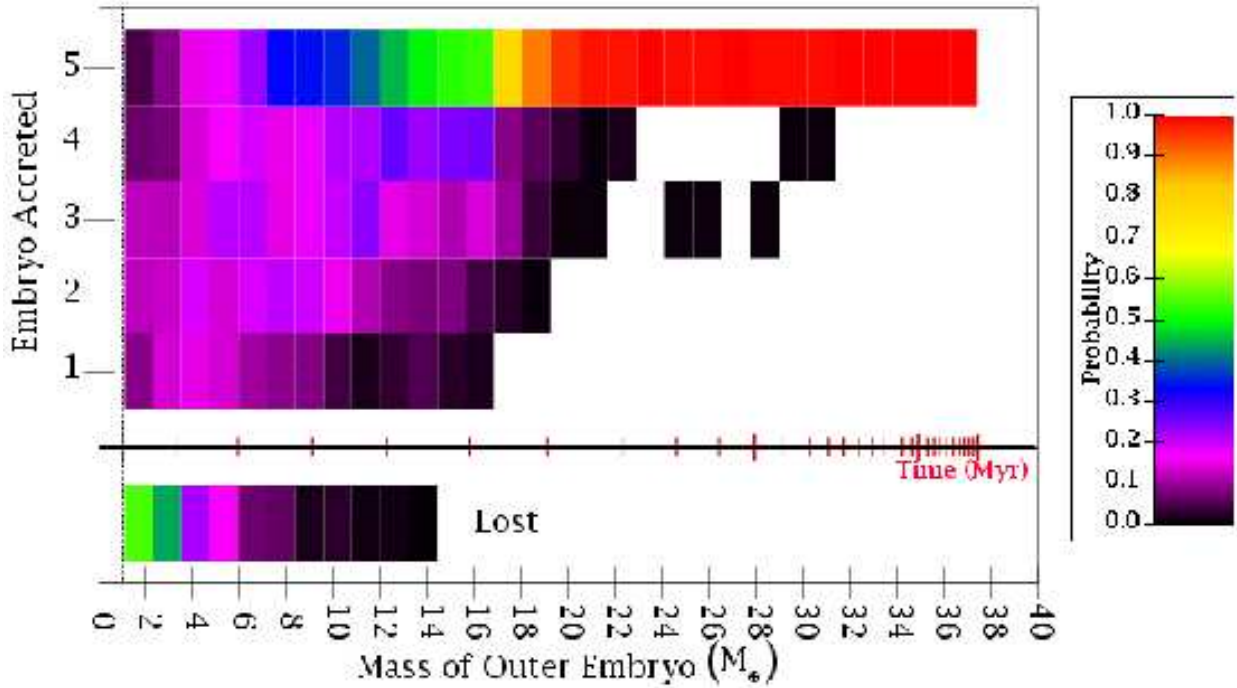}
%\begin{figure}[h!]
%\vglue 3.5truein
%\special{psfile=PS/prob.ps angle=0 hoffset=-10 voffset=-180 vscale=80 hscale=80}
\caption{\footnotesize \label{fig:prob} The probability that a
fragment created beyond the orbits of the embryos will suffer a
particular fate as a function of the mass of the outermost embryo.
In this figure, the color represents the probability of the fate
given on the ordinate. A fragment can either be accreted by an
embryo (which are numbered as a function of increasing heliocentric
distance), or get too close to the Sun (indicated by 'Lost'). The
red tickmarks indicate time, where the small and large marks show a
spacing of $10^5$, and $10^6$ years, respectively.}
\end{figure}

The dichotomy between the outer embryo and the rest of the oligarchs
supplies an important insight into how the accretion efficiency varies
in these simulations. Clearly, near the end of the simulation, the
outer embryo gets large enough that fragments generated in the outer
disk (recall that the outer embryo moves between 6.5 and $8.5\,$AU,
while the disk extends to $16\,$AU) cannot escape as they sweep by.
Thus, it grows, while its neighbors starve. \Red{This} situation was
not always the case. Fig$.$~\ref{fig:prob} shows the fate of fragments
that were created beyond the orbits of the growing cores as a function
of the outer embryo mass. In particular, the ordinate indicates the
fate of a fragment, which can be either accretion by Embryo~$N$, where
$N$ ranges from 1 to 5 in order of increasing heliocentric distance,
or getting too close to the Sun (marked as `Lost'). The abscissa is
the mass of the outer embryo. Note that at early times (i$.$e$.$
before $400,000$ years) this can be used as a proxy for the mass of
all the embryos since they were roughly the same at these times
(Fig$.$~\ref{fig:raf}B). The color in the figure shows the probability
that a fragment that formed far from the Sun will reach one of the
fates listed.

When the cores were $\sim\!1\,M_{\oplus}$, the \Red{probability} that
an embryo will capture a fragment as it sweeps by is small and thus
the embryos grow at an equal rate. Note that at this time, half of
these fragments were removed from the simulation because they got too
close to the Sun, implying that a particle only had a 13\% chance of
being accreted by any individual embryo. This efficiency is similar to
that found by Kary et al$.$~(1993) for a core with relatively small
envelope. By the time the embryos reach $10\,M_\oplus$, the
probability that a particle will survive the passage of an individual
embryo is roughly 40\% --- reaching 90\% at a mass of $18\,M_\oplus$.
\Red{This} higher efficiency is due to the fact that an extended gas
envelope has caused the effective radius of capture of fragments to be
a fairly large fraction of the embryo's Hill sphere. \Red{It} is
important to note that our simulations do not include run-away gas
accretion or allow the embryos to open a gap in the gas disk. Thus,
these simulations are not applicable for masses larger than
$\sim\!10\,M_\oplus$.

\begin{figure}
\epsscale{1.}
\plotone{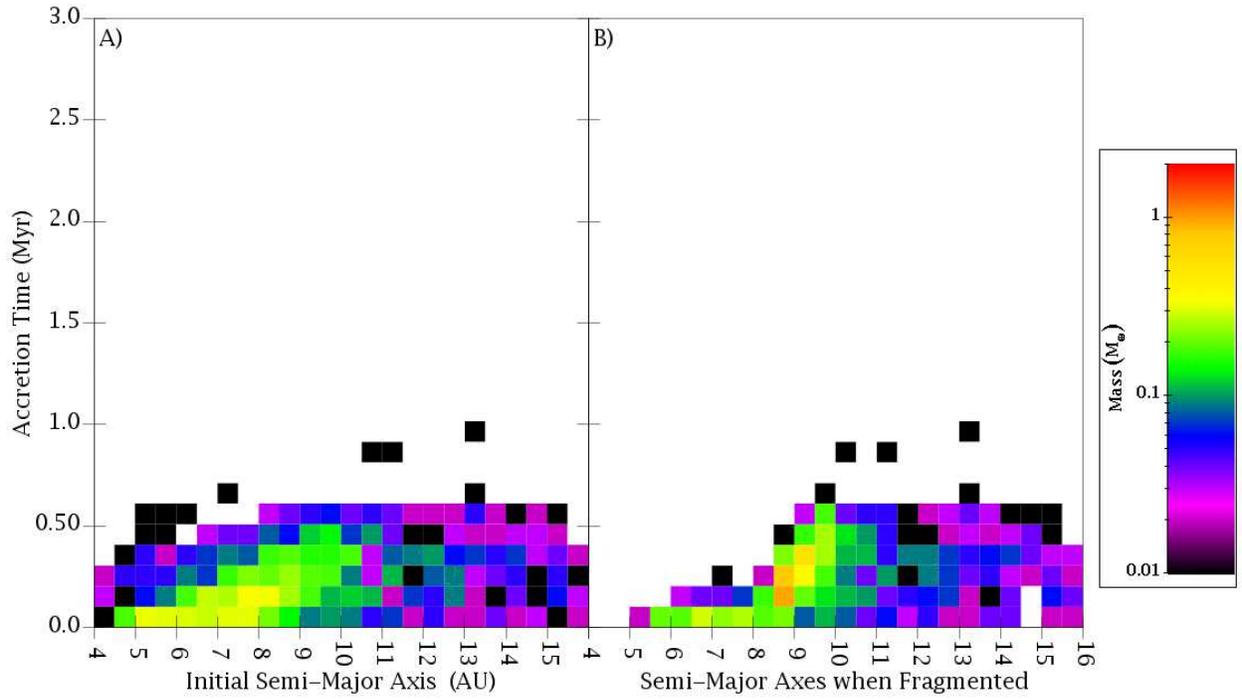}
%\begin{figure}[h!]
%\vglue 3.7truein
%\special{psfile=PS/atM_raf4.ps angle=0 hoffset=-10 voffset=-20 vscale=45 hscale=45}
\caption{\footnotesize \label{fig:raf4} An analysis of objects
eaten by the Embryo~3 of the simulation shown in
Fig$.$~\ref{fig:raf}. In particular, the color shows the amount of
mass accreted as a function of location (on the abscissa) and the
time that the object was accreted (ordinate). A) The initial
semi-major axis of the planetesimal. B) The Semi-major axes where
the planetesimal fragmented.}
\end{figure}

Given the above discussion, it might be tempting to conclude that most
of the mass accreted by the inner four embryos came from the distant
disk. However, this is not the case. Fig$.$~\ref{fig:raf4} shows
characteristics of the particles that were accreted by the Embryo~3
during the simulation. \Red{The} abscissa in Fig$.$~\ref{fig:raf4}A is
the semi-major axes of a particle at the beginning of the simulation
\Red{and} the ordinate is the time that the particles \Red{were}
accreted.  The color is the total mass from that particular semi-major
axes that was accreted at that particular time.
Fig$.$~\ref{fig:raf4}B is similar, but instead of showing the initial
semi-major axes, we plot the semi-major when the particle became a
fragment because of a collision.

We interpret this figure as follows. In Fig$.$~\ref{fig:raf4}A, the
accretion occurred during early times (which we already know from
Fig$.$~\ref{fig:raf}). In addition, the whole region between 4 and
$8\,$AU is yellow, indicating that the core mainly feeds on objects
that were basically uniformly spread between the embryos. However,
Fig$.$~\ref{fig:raf4}B shows that most of these particles accreted by
this embryo suffered catastrophic collisions near $\sim\!9\,$AU. This
is due to the fact that, although the objects eaten by Embryo~3 formed
in amongst the embryos, they were initially scattered outward forming
a dense ring of material outside of the orbits of the embryos. This is
similar to what happened in Run~A (c.f$.$ Fig$.$~\ref{fig:run0}).
\Red{Once} in this ring \Red{the planetsimals fragmented} and spiraled
inward due their increased drag, only to impact an embryo. Once the
outer embryo became large enough to capture all of the fragments,
Embryo~3 stopped growing. Indeed, objects that were accreted by the
outermost embryo show a similar behavior (see Fig$.$~\ref{fig:raf6})
in that they were scattered into the ring before they fragmented.
However, this embryo continued to feed for the length of the
simulation because of its favorable location.

\begin{figure}
\epsscale{1.}
\plotone{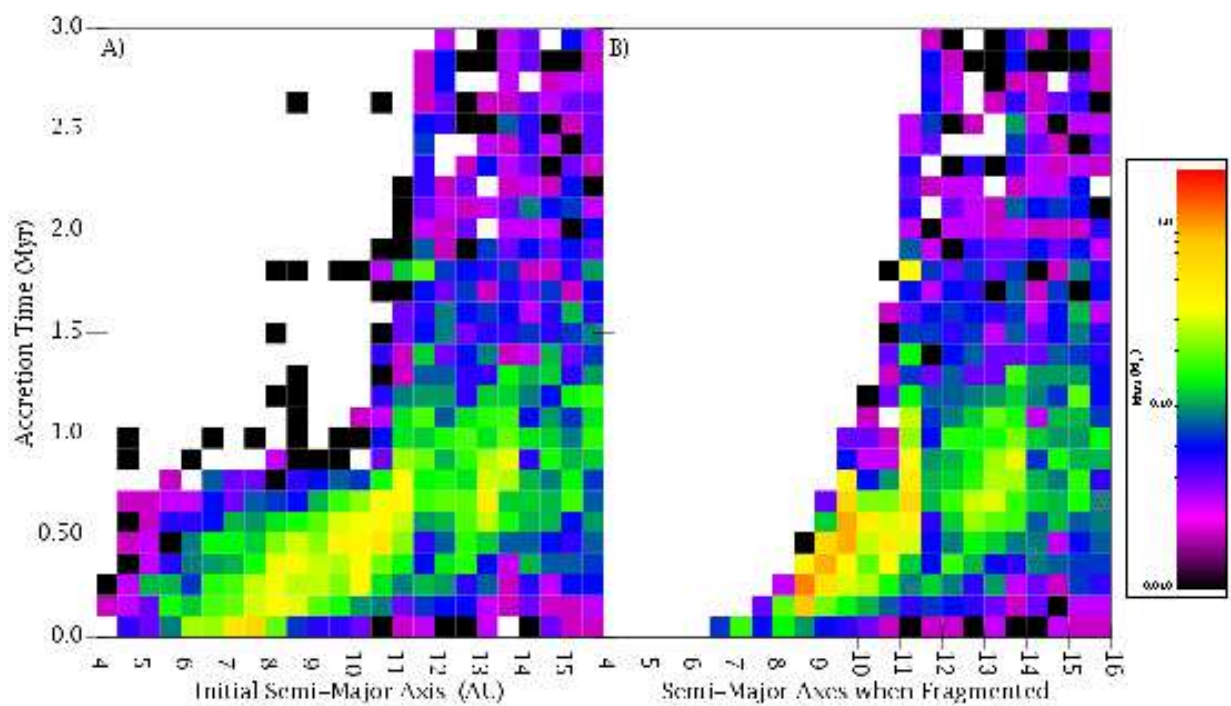}
%\begin{figure}[h!]
%\vglue 3.4truein
%\special{psfile=PS/atM_raf6.ps angle=0 hoffset=-10 voffset=-200 vscale=80 hscale=80}
\caption{\footnotesize \label{fig:raf6} Same as Fig$.$~\ref{fig:raf4},
but for the outermost embryo.}
\end{figure}

With the insight gleamed from the above discussion, we can now use
Fig$.$~\ref{fig:prob} to grasp why we only get significant growth when
$r_{\rm f}\!=\!10\,$m. As described above, this figure shows that at
the beginning of the simulation, the capture efficiency of the embryos
is about 13\%. While this is small, it is large enough to allow the
embryos to begin to grow. \Red{This is illustrated in a run we
  performed that is similar} to Run~C, but with $r_{\rm f}\!=\!3\,$m.
For this simulation, the capture efficiency was only 2.7\% (again
consistent with the results of Kary et al$.$ for such rapidly
migrating fragments) and thus although $95\,M_\oplus$ of fragments
went through the region \Red{populated by} the embryos, the embryos
grow to only $\sim\!3.6\,M_\oplus$. For the same $95\,M_\oplus$
\Red{of fragments}, an accretion probability of 13\% predicts that the
embryos should grow to $12\,M_\oplus$ each. Thus, the capture
efficiency is simply too small for $r_{\rm f}\!<\!10\,$m. And since
the fragments force the embryos too close to the Sun for $r_{\rm
  f}\!>\!10\,$m, we must conclude that significant growth only occurs
for a narrow range of aerodynamic drag parameters.

The other issue along these lines that must be discussed is the role
of the embryo atmospheres. Here, luckily we have some good news. We
performed a series of runs similar to Run~C, but where we varied
\Red{the opacity of the atmosphere, $\kappa$,} from 0.002 to 100 times
the interstellar value. We find that significant growth in all the
simulations, although the resulting embryos were slightly less massive
in the simulations where $\kappa$ was larger (i$.$e$.$ the effective
radius of the embryos were smaller, see Eq$.$~\ref{eq:rc}). Note,
however, that the embryos only grew to $\sim\!3\,M_\oplus$ in the
simulation with no atmosphere at all.  \Red{These results demonstrate
  that} although an atmosphere is required for significant growth, its
usefulness is not particularly sensitive to opacity.

Earlier in this subsection we mentioned that we performed simulations
with $r_{\rm p}\!=\!10\,$km in the hope that planetesimal driven
outward migration would counteract the tendency for the fragments to
push the embryos inward. To test this hypothesis, we performed seven
simulations like Run~C, but with $r_{\rm p}\!=\!10\,$km and $r_{\rm
f}\!=\!10\,$m. The only difference between the runs is that we used
a different random number seed and moved the outer embryo by roughly
$10^{-5}\,$AU. In all cases, the embryos were pushed inward and out
of our simulation. Thus, we tentatively conclude that planetesimal
driven outward migration in not an important process in situations
where push-in occurs.

Note that the embryos do migrate outward in simulations that have
fragmentation if $r_{\rm f}$ is small enough that resonant trapping
does not occur. However, they do not grow as massive in these
simulations as they did when fragmentation was turned off. \Red{In} a
simulations with $r_{\rm p}\!=\!10\,$km and $r_{\rm f}\!=\!1\,$m, the
outer embryo only grew to $13\,M_\oplus$. Note that this is still
large enough to be considered a legitimate giant planet core.
\Red{The} only way we see to save planetesimals driven migration as a
viable mechanism for core accretion is if either: 1) the planetesimals
are strong so that they do not break, 2) they are pulverized during
collisions so that all the fragments are smaller than $\sim\!10$m, or
3) they undergo a collisional cascade fast enough that they are not
dynamically important when they are $\sim\!100$m in size.

We now \Red{return} to the issue of whether fragmentation can be a
general aid to growth and ask if there is a way to circumvent the
problem that substantial growth only apparently occurs for a narrow
range of aerodynamic drag parameters. One possibility is to once again
invoke our $\sim$Mars-mass embryos. It is possible to imagine a
situation where these smaller embryos gravitationally scatter
fragments out of resonances, thereby stopping their inward push on the
embryos and allowing the fragments to be accreted.  \Red{This} turns
out not to be the case. We performed three simulations similar to the
ones shown in Fig$.$~\ref{fig:mars}, but where we turned on
fragmentation with $r_{\rm f}\!=\!10\,$m.  Unfortunately, in all the
cases, the original Earth-mass embryos were pushed out of the
simulated region by the fragments.

What other alternatives are there? As described several places above,
once a tracer becomes a fragment of a particular $r_{\rm f}$, we did
not allow it to collisionally evolve any further. Although, we argue
above that collisions will not significantly alter the dynamics of the
resonant particles, it is possible that collisions will fragment the
objects further so that they will be able to leave the resonance due
to aerodynamic drag. This may be important given that the collisional
optical depth can be much larger than one. We \Red{can only} speculate
as to how the fragmentation of our fragments will affect the above
results. \Red{Some} insight is available in the literature. We find
that the typical impact velocity between objects in the resonances is
roughly $100\,$m/s. For objects of, say, $100\,$m the critical speed
for catastrophic disruption is only $3\,$m/s (Stewart \&
Leinhardt~2009). These collisions are so \Red{energetic} that the
resulting fragments are probably so small (Stewart-Mukhopadhyay,
pres$.$ comm.) that they would sweep by the planets without getting
accreted.

More generally, our results reveal that accretion is only efficient
for a narrow range of planetesimal sizes, and, although we model the
collisional cascade \Red{crudely}, we see no reason why this result
should not extend to more realistic situations. Both observations of
asteroid families (see Zappal{\` a} et al$.$~2002 for a review) and
analytic and numerical models (BA99; Durda et al$.$~2007) show that
collisionally processed populations exhibit a size distribution that
approximately follows a power-law.  Thus, we believe that if we could
\Red{model} the collisional evolution of our system more accurately,
we would find that only a small fraction of the small bodies would be
found in the `sweet-spot' for accretion.  As a result, we believe that
fragmentation is not the general solution to the core-formation
problem.

\subsection{The Role of Evaporation and Condensation}
\label{ssec:aero}

As we described in \S{\ref{sec:review}}, the problem of core formation
might be solved if models took into account the increase of the solid
surface density due to the interplay of the evaporation of icy
planetesimals that moved interior to the snow line and the
recondensation of water from the diffusion of vapor back outward
(CZ04). \Red{CZ04's} models suggest that the solid surface density
could be enhanced by more than an order of magnitude in the region
between $\varpi_{\rm SL}$ and $\varpi_{\rm SL}\!+\!d\varpi_{\rm
  diff}$, where $\varpi_{\rm SL}$ is the location of the snow-line and
$d\varpi_{\rm diff}$ is the diffusion length scale. This material
could be used to grow the giant planet cores.

In this subsection, we test the above hypothesis with direct $N$-body
calculations using a modified version of the code described in
\S{\ref{sec:code}}. In order \Red{to} simply mimic evaporation and
recondensation, this code removes any tracer particle \Red{that}
evolves to a heliocentric distance less than $\varpi_{\rm SL}$ and
replaces it with a new tracer with a heliocentric distance ($\varpi$)
chosen at random from a uniform distribution between $\varpi_{\rm SL}$
and $\varpi_{\rm SL}\!+\!d\varpi_{\rm diff}$. Note that we are
overestimating the outward shift of material because physically this
is a diffusion process which puts more material near $\varpi_{\rm SL}$
than our uniform distribution does. We set $\varpi_{\rm
  SL}\!=\!3.9\,$AU, which is slightly interior to the initial inner
edge of our planetesimal disk at $4\,$AU, and $d\varpi_{\rm diff}
\!=\!1\,$AU, in accord with the results of CZ04's modeling. Since this
new tracer represents objects condensing from the gas disk, we place
it on a circular orbit with an inclination of
$\frac{1}{2}\tan{(z_s/\varpi)}$.

\Red{Once} a planetesimal evaporates, not all of its mass diffuses
back to outside the snow-line and recondenses. \Red{To account for
  this} we set the mass of the new tracer to $\epsilon$ times the mass
of the original.  The way to interpret this is to recall that each
tracer particle actually represents $N$ objects of radius $r_{\rm p}$
or $r_{\rm f}$, depending on whether they are `planetesimals' or
`fragments'. By changing the mass of the tracer, we are effectively
decreasing $N$.  The values of $r_{\rm p}$ and $r_{\rm f}$ remain
fixed during the simulation. For lack of a better constraint, we used
$\epsilon\!=\!0.75$ for our original series of runs. This is a
conservative value (in the sense of promoting embryo growth) given
that the original planetesimals in this region are probably less than
half water ice.

We performed five simulations with $r_{\rm p}\!=\!10\,$km and $r_{\rm
f}\!=\!1$, 3, 10, 30, and $100\,$km. All other free parameters are
the same as in Run~A. Our results show trends very similar to the
runs without evaporation. In the runs with $r_{\rm f}\!\geq\!30\,$m,
fragments build up in the embryos' resonances, and the embryos are
pushed inward. Unlike the previous runs where the embryos were pushed
out of the simulation region, however, here the embryos migrated until
most of their strong mean motion resonances were closer to the Sun
than the snow-line. For example, the runs with $r_{\rm f}\!=\!100\,$m
and $30\,$m migrated so that the heliocentric distance of the 2:1 and
5:3, respectively, were slightly less than $\varpi_{\rm SL}$, i$.$e$.$
their semi-major axes were at 2.5 and $2.8\,$AU, respectively. Once
this occurred, the embryos stopped growing. The most massive embryo in
these simulations was only $7.1\,M_\oplus$.

In the runs $r_{\rm f}\!\leq\!3\,$m, the fragments are small enough
that they sweep by the planets. Recall that there is very little
growth in similar runs without evaporation. Here, only the inner
embryo grows because it sits in the condensation region. For both of
the runs that showed this behavior ($r_{\rm f}\!=\!1\,$m and $r_{\rm
f}\!=\!3\,$m), the inner embryo grew to $10\,M_\oplus$ in 3 million
years. However, again only one core grew and thus these systems are
not like the Solar System. In addition, even with
evaporation/recondensation, the systems lost far more mass
($\sim\!90\,M_\oplus$) to the inner regions than accreted onto the
planet. This indicates that $\epsilon$ is a critical factor in these
simulations and recall that we used a conservative value for this. In
a simulation with $\epsilon\!=\!0.25$, the inner embryo had a mass of
only $4\,M_\oplus$ at the end of the calculation. This is probably
more typical of what we should expect in more realistic situations.

\begin{figure}
\epsscale{1.}
\plotone{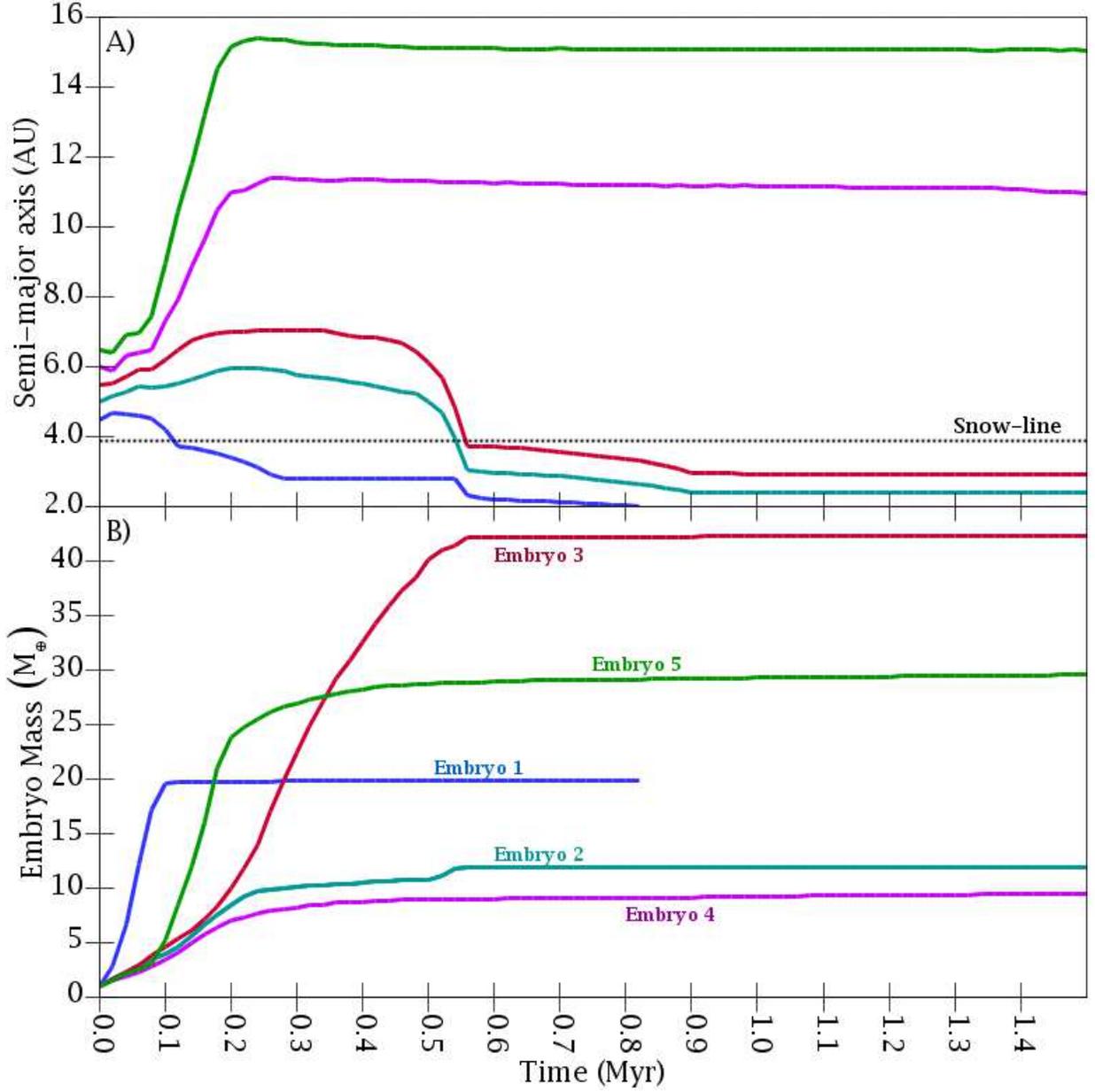}
%\begin{figure}[h!]
%\vglue 6.6truein
%\special{psfile=PS/aM_evap.ps angle=0 hoffset=0 voffset=10 vscale=60 hscale=60}
\caption{\footnotesize \label{fig:evap} The temporal evolution of the
  five embryos in the evaporation run with $r_{\rm f}\!=\!10\,$m,
  $r_{\rm p}\!=\!10\,$km, and $\epsilon=0.75$. A) The semi-major axes
  as a function of time. Each embryo is represented by a solid curve
  of a different color. The dotted line shows the assumed location of
  the snow-line. Embryo~1 falls out of the region that we simulate at
  $830,000$ years. This is a numerical artifact, however. B) Embryo
  mass as a function of time. The color solid curves corresponds to
  the semi-major axes in (A).}
\end{figure}

Once again interesting things occur at \Red{$r_{\rm f}\!=\!10\,$m}. In
this run, the evolution of which is shown in Fig$.$~\ref{fig:evap}, we
see many of the processes thus far discussed combine to form four
large cores, and thus we now describe its evolution in detail. For the
first $100,000$ years, the embryos grow and spread. Fragments are
being accreted mainly by the innermost embryo (Embryo~1) because it
sits in the condensation zone and it grows to $19\,M_\oplus$. Recall
that we do not allow the embryos to accrete nebula gas or open gaps.
Thus, from this time onward our simulation is missing physics that
would potentially dominate the evolution. Having said this, the
subsequent evolution supplies important insight into how the various
processes interact, and so, we continue our description.

Between $80,000$ and $230,000$ years self-sustaining planetesimal
migration causes the outer four embryos to migrate outward. During
this time, the outermost (Embryo~5) grows to $26\,M_\oplus$. As the
embryos spread, fragments start to build up in the mean motion
resonances of the Embryo~1. We believe that this occurs simply
because the outward migration creates enough room between the two
inner embryos to allow fragments to settle there. Embryo~1 starts to
migrate inward and its growth stops. By 280,000 years its 5:3 mean
motion resonance moves inside of the snow-line and it also stops
migrating.

At this point, the action moves to the region between 8 and $11\,$AU.
The self-sustaining outward migration left this region devoid of
embryos and thus planetesimals start to concentrate there. As this
occurs their collision rate is large and many fragments are formed.
The fragments spiral in and are accreted by Embryo~3. Between
$230,000$ and $460,000$ years, this embryo grows from 14 to
$38\,M_\oplus$. It is only at the end of this time that the embryo
grows sufficiently large to trap the fragments in its exterior
resonances. These fragments eventually push the inner three embryos
toward the Sun. At $830,000$ years Embryo~1 was pushed out of our
simulation region. It had a mass of $20\,M_\oplus$. The outer two
embryos did not get pushed around by the fragments because their
outward migration depleted the outer disk enough that the collision
rate was small and so fragments did not form. At the end of the
simulation there were four cores with masses ranging from 9 to
$42\,M_\oplus$. Presumably, the original Embryo~1 would also be
present if we were not forced to remove it for technical reasons, as
explained above.

We can only speculate as to how the system would have evolved if we
had allowed the embryos to directly accrete gas. All would have
proceeded as before until Embryo~1 reached $\sim\!10\,M_\oplus$ at
60,000 years. We believe that this growth would not have stopped the
onset of planetesimal driven migration, however. \Red{If} it had any
effect at all, it would have \Red{triggered} the migration at an
earlier time.  Thus, Embryo~6 would have moved outward and also grown
to the point where it would have accreted gas. The embryo reached
$10\,M_\oplus$ at 130,000 years in our simulation; at which point is
was at $11\,$AU. It is unclear how the system would have behaved after
this time. However, we believe that it safe to conclude that we would
have created two gas giant planets in this system --- and in only
130,000 years. All this thanks to a combination of
evaporation/recondensation and planetesimal driven migration.

\section{Summary \& Conclusions}
\label{sec:concl}

We presented the results of a large number of $N$-body simulations of
the formation of giant planet cores. Our goal was to measure the
effectiveness of several mechanisms for enhancing overall growth
rates. To achieve this, for most of our simulations we started with a
simple system containing 5 Earth-mass embryos embedded near the inner
edge of a planetesimal disk. This disk contained $200\,M_\oplus$ of
material that was spread from 4 to $16\,$AU. All the planetesimals
were assumed to have the same radius $r_{\rm p}$. Our simulations
potentially included a combination of the following processes: 1)
aerodynamic drag on the small-bodies, 2) collisional damping, 3)
extended atmosphere around the embryos (Inaba and Ikoma~2003), 4)
embryo eccentricity damping due to gravitational interaction with the
gas disk, 5) fragmentation of the planetesimals to form objects with
radius $r_{\rm f}$, and 6) evaporation and recondensation at the snow
line (Cuzzi \& Zanhle~2004). Not all simulations employed all these
processes. Our goal was to determine which of these processes would
allow several $10\,M_\oplus$ cores to form in less than 3 million
years. Unless noted, these simulation assumed a gas disk with a
surface density of 5 times the minimum mass solar nebula value at 5AU,
and varying in inverse proportion to heliocentric distance.

We first presented an example of the the most basic simulations we
performed (Run~A). In particular, it included Processes~(1)--(4), but
not (5) or (6). During this calculation the embryos quickly scattered
planetesimals out of the region that they inhabited --- forming two
massive rings immediately adjacent to this region (see
Fig$.$~\ref{fig:run0}). The planetesimals became decoupled from the
embryos due to the effects of aerodynamic drag. Very little growth
occurred. This run illustrates the first major conclusion of this work
--- the gravitational interaction between the embryos and the
planetesimals lead to the wholesale redistribution of material.
Therefore, they must be handled in a realistic manner in order to
produce reasonable results. This conclusion is true for all of our
simulations.

Run~A is representative of 90\% of our simulations without
fragmentation. The remaining 10\%, however, exhibit quite dramatic
behavior. In these simulations, we see a burst of outward migration.
The outer embryos race through the distant disk, eating as they go.
On timescales of the order of $120,000$ years, the outer embryo can
migrate $\sim\!6\,$AU and grow to roughly $30\,M_\oplus$. In most of
these simulations, several of the five embryos move outward and grow
significantly. Indeed, in the example shown in Fig$.$~\ref{fig:rawayf}
the cores have masses of between 3.2 and $29\,M_\oplus$ by the end of
the simulation.

Despite the success we have had with this so-called {\it planetesimal
  driven migration}, there is a problem --- it occurred in only 10\%
of our simulations. However, we performed a set of simulations where
we introduced 10 $\sim$Mars-mass embryos into the system. Although,
there are some issues with these calculations (see
\S{\ref{ssec:aero}}), we see planetesimal driven migration in all
cases. Thus, we believe that if \Red{we} had been able to include a
more realistic mass-distribution for the embryos, more of our systems
would have exhibited this behavior.

We next performed a set of simulations which included the
fragmentation of planetesimals. It was hypothesized that the creation
of fragments would substantially increase embryo growth rates because
the fragments' inclinations \Red{would quickly} damp (due to their
small sizes) and they would be in the shear-dominated regime when they
encountered the embryos (Wetherill and Stewart 1993; R04).  However,
we found that this mechanism promotes rapid embryo growth only for a
narrow range of parameters. In particular, for our nebular model,
$r_{\rm f}\!\leq\!3\,$m fragments were moving so quickly that they
stream by the embryos without being accreted. For $r_{\rm
  f}\!\geq\!30\,$m, the fragments pile up in mean motion resonances
with the embryos rather than being accreted. Indeed, enough material
gets trapped in the resonances that it can push the embryos into the
inner Solar System (see Fig$.$~\ref{fig:push}).

Only for $r_{\rm f}\!=\!10\,$m do we see a significant amount of
growth. Indeed, for this narrow set of parameters, we produce cores
larger than $10\,M_\oplus$ by our 3 million year time limit.
Unfortunately, we were unable to find a way to broaden the range of
parameters over which this mechanism functions. Therefore, we are not
very optimistic that this mechanism will prove valuable in the long
run. Indeed, in most cases it did more harm than good.

Of particular concern is the fact that the fragments can get trapped
in the mean motion resonances with the embryos and then push them
around. We believe that our simulations show that icy planetesimals
must break in such a way to avoid putting much mass in fragments with
radii between $\sim\!30\,$m and $\sim\!1\,$km. Fortunately, there is
observational evidence that supports this conclusion. In particular,
we note that of the $\geq\!50$ asteroid families thus far discovered
(Nesvorn{\' y} et al$.$~2006), none are composed of what are believed
to be the most comet-like asteroid types (P and D-types).  This
despite the fact that P and D-types make up at least 10\% of known
asteroids (Moth{\'e}-Diniz et al$.$~2003). Indeed, although
$\sim\!90\%$ of the Trojans asteroids are P/D-types, the only asteroid
family in the swarms is a more normal C-type (M.~Bro{\v z}, pers.
comm.). The simplest interpretation for these observations is that
when a comet-like object breaks apart, it is pulverized to small
sizes. This is also supported by crater counts on the Galilean
satellites, which show a dearth of primary impactors in the size range
we are discussing (Bierhaus et al$.$~2005).

Finally, we did a set of simulations that modelled the evaporation of
planetesimals at the snow-line (which we put at $\sim\!4\,$AU) and had
them to recondense at random locations between $4\,$AU and $5\,$AU.
This process will increase the surface density of solids in this
region and thereby increase accretion rates. We find that in
simulations with fragments radii greater then $30\,$m, the embryos are
pushed inside the snow-line by resonant-trapped fragments and very
little growth occurs. For $r_{\rm f}\!\leq\!3\,$m, the fragments
sweep by all but the innermost embryo. As a result, only the inner
embryo grows. Although this can produce a core of $10\,M_\oplus$, it
is an inefficient process, and 90\% of the fragments are lost to the
inner Solar System. We performed one simulation with $r_{\rm
f}\!=\!10\,$m which produced a system with many cores more massive
than $10\,M_\oplus$. In this simulation, the growth of the innermost
core triggered planetesimal driven migration in the rest of the
embryos (see Fig$.$~\ref{fig:evap}).

In summary, we find that widely used approximations for core accretion
typically overestimate the ability of models to produce sufficiently
large cores because they do not incorporate the dynamical influences
which redistribute planetesimals near the feeding zones of the
embryos.  As a result of this, we believe that planetesimal driven
outward migration offers the best hope for solving the issue of giant
planet core formation. \Red{Clearly} more work needs to be done to
demonstrate it effectiveness.  In particular, the effects of a
realistic size distribution must be taken into account. This is
computationally challenging and thus we leave it for future work.

\acknowledgments This work as been directly supported by a grant from
the National Science Foundation (Award ID 0708775). HFL is also
grateful for funding from NASA's Origins, and OPR programs.  We would
like to thank Bill Bottke, John Chambers, and Alessandro Morbidelli
for useful discussions.

\clearpage

\section*{References}

\begin{itemize}
  \setlength{\itemindent}{-30pt} \setlength{\labelwidth}{0pt}

\item[] Adachi, I., Hayashi, C., and Nakazawa, K.: 1976,
{\it Progress of Theoretical Physics} {\bf 56}, 1756.

\item[] Agnor, C.B., Canup, R.M., and Levison, H.F.: 1999,
{\it Icarus} {\bf 142}, 219.

\item[] Alibert, Y., Mordasini, C., Benz, W., and
Winisdoerffer, C.: 2005, {\it Astron.  Astroph.} {\bf 434}, 343.

\item[] Benz, W., Asphaug, E.\ 1999.\ Catastrophic Disruptions
  Revisited.\ {\it Icarus} {\bf  142}, 5-20.

\item[] Bierhaus, E.~B., 
Chapman, C.~R., Merline, W.~J.\ 2005.\ Secondary craters on Europa and 
implications for cratered surfaces.\ {\it Nature} {\bf 437}, 1125-1127. 

\item[] Brasser, R., Duncan, 
M.~J., Levison, H.~F.\ 2006.\ Embedded star clusters and the formation of 
the Oort Cloud.\ {\it Icarus} {\bf  184}, 59-82. 

\item[] Capobianco, C., Duncan, M., and Levison, H.: 2009 in preparation

\item[] Chambers, J.\ 2008.\ 
Oligarchic growth with migration and fragmentation.\ {\it Icarus} {\bf 198}, 256-273.

\item[] Chambers, J.E.~and Wetherill, G.W.: 1998, {\it
Icarus} {\bf 136}, 304.

\item[] Chambers, J.E.: 2001, {\it Icarus} {\bf 152}, 205.

\item[] Cuzzi, J.N.~and Zahnle, K.J.: 2004, {\it
Astrophys. J.} {\bf 614}, 490.

\item[] Durda, D.~D., Bottke, W.~F., Nesvorn{\'y}, D., Enke, B.~L.,
  Merline, W.~J., Asphaug, E., Richardson, D.~C.\ 2007.\ Size
  frequency distributions of fragments from SPH/N-body simulations of
  asteroid impacts: Comparison with observed asteroid families.\ 
  Icarus 186, 498-516.

\item[] Duncan, M.J., Levison, H.F., and Lee, M.H.: 1998,
{\it Astronomical Journal} {\bf 116}, 2067.

\item[] Fernandez J.~A$.$ and Ip, W.-H.: 1984, {\it Icarus} 
{\bf 58}, 109.

\item[] Goldreich, P., Lithwick, Y., and Sari, R.: 2004a,
{\it Annual Review of Astronomy and Astrophysics} {\bf 42}, 549.

\item[] Goldreich, P., 
Lithwick, Y., Sari, R.\ 2004b.\ Final Stages of Planet Formation.\ 
{\it Astrophysical Journal} {\bf 614}, 497-507.

\item[] Gomes, R.S., Morbidelli, A., and Levison, H.F.:
2004, {\it Icarus} {\bf 170}, 492.

\item[] Hahn, J.M.~and Malhotra, R.: 1999, {\it
Astronomical Journal} {\bf 117}, 3041.

\item[] Haisch, K.E., Jr., Lada, E.A., and Lada, C.J.:
2001, {\it Astrophys. J.} {\bf  553}, L153.

\item[] Halliday, A.: 2004, {\it Nature} {\bf 431}, 253.

\item[] Hayashi, C.: 1981, {\it Prog. Theor. Phys.}
{\bf  70}, 35. 

\item[] Hayashi, C., Nakazawa, K., Nakagawa, Y.\ 1985.\ Formation of
  the solar system.\ {\it Protostars and Planets II} 1100-1153.
  
\item[] \Red{Hillenbrand, L.~A.\ 2008, Physica Scripta Volume T, 130,
    014024}

\item[] Hubickyj, O., Bodenheimer, P., and Lissauer, J.J.:
2005, {\it Icarus} {\bf 179}, 415.

\item[] Ida, S.~and Makino, J.: 1993, {\it Icarus} {\bf
106}, 210.

\item[] Inaba, S.~and Wetherill, G.W.: 2001, {\it Lunar and
Planetary Institute Conference Abstracts} {\bf 32}, 1384.

\item[] Inaba, S.~and Ikoma, M.: 2003, {\it
Astron. Astroph.} {\bf 410}, 711.

\item[] Inaba, S., Wetherill, G.W., and Ikoma, M.: 2003,
{\it Icarus} {\bf 166}, 46.

\item[] Kary, D.~M., Lissauer, 
J.~J., Greenzweig, Y.\ 1993.\ Nebular gas drag and planetary accretion.\ 
{\it Icarus} {\bf 106}, 288.

\item[] Kirsh, D.~R., Duncan, M., 
Brasser, R., Levison, H.~F.\ 2009.\ Simulations of planet migration driven 
by planetesimal scattering.\ {\it Icarus}{\bf 199}, 197-209. 

\item[] Kokubo, E.~and Ida, S.: 1998, {\it Icarus} {\bf
131}, 171.

\item[] Kokubo, E.~and Ida, S.: 2000, {\it Icarus} {\bf
143}, 15.

\item[] Kominami, J., Tanaka, H., and Ida, S.: 2005, {\it
Icarus}  {\bf 178}, 540.

\item[] Korycansky, D.G.~and Pollack, J.B.: 1993, {\it
Icarus}  {\bf 102}, 150.

\item[] Laughlin, G.~and Adams, F.C.: 1997, {\it
Astrophys. J.} {\bf 491}, L51.

\item[] Levison, H.~F., 
Duncan, M.~J.\ 2000.\ Symplectically Integrating Close Encounters with the 
Sun.\ {\it Astronomical Journal} {\bf 120}, 2117-2123. 

\item[]  Levison, H.~F., 
Morbidelli, A.\ 2007.\ Models of the collisional damping scenario for 
ice-giant planets and Kuiper belt formation.\ {\it Icarus} {\bf 189}, 196-212.

\item[] Levison, H.~F., 
Morbidelli, A., Gomes, R., Backman, D.\ 2007.\ Planet Migration in 
Planetesimal Disks.\ {\it Protostars and Planets V} 669-684. 

\item[] Lissauer, J.~J.\ 1987.\ 
Timescales for planetary accretion and the structure of the protoplanetary 
disk.\ {\it Icarus} {\bf  69}, 249-265. 

\item[] McNeil, D., Duncan, M., and Levison, H.F.: 2005,
{\it Astronomical Journal}  {\bf 130}, 2884.

\item[] Mizuno, H.: 1980, {\it Progress of Theoretical
Physics}  {\bf 64}, 544.

\item[] Mizuno, H., Nakazawa, K. and Hayashi, C.:
1978. {\it Prog. Theor. Phys.}  {\bf 60}, 699.

\item[] Moth{\'e}-Diniz, T., Carvano, J.~M., \& Lazzaro, D. 2003.
  Distribution of taxonomic classes in the main belt of asteroids.\ 
  Icarus 162, 10-21.

\item[] Nesvorn{\'y}, D., Bottke, W.~F., Vokrouhlick{\'y}, D.,
  Morbidelli, A., Jedicke, R.\ 2006.\ Asteroid families.\ Asteroids,
  Comets, Meteors 229, 289-299.

\item[] Paardekooper, S.-J., Papaloizou, J.~C.~B.\ 2008.\ On disc
  protoplanet interactions in a non-barotropic disc with thermal
  diffusion.\ {\it Astronomy and Astrophysics} {\bf 485}, 877-895.
 
\item[] Papaloizou, J.C.B.~and Larwood, J.D.: 2000, {\it
Monthly Notices of the Royal Astronomical Society}  {\bf 315}, 823.

\item[] Pollack, J.B., Hubickyj, O., Bodenheimer, P.,
Lissauer, J.J., Podolak, M., and Greenzweig, Y.: 1996, {\it Icarus}
{\bf 124}, 62.

\item[] Rafikov, R.R.: 2004, {\it Astronomical Journal}
{\bf 128}, 1348.

\item[] Raymond, S.~N., Scalo, 
J., Meadows, V.~S.\ 2007.\ A Decreased Probability of Habitable Planet 
Formation around Low-Mass Stars.\ {\it Astrophysical Journal} {\bf 669}, 606-614.

\item[] Skeel, R.~D., \& Biesiadecki, J.~J.: 1994, {\it
Ann.\ Numer.\ Math.}  {\bf 1}, 191.

\item[] Stevenson, D.J.~and Lunine, J.I.: 1988, {\it
Icarus} {\bf 75}, 146.

\item[] Touboul, M., Kleine, 
T., Bourdon, B., Palme, H., Wieler, R.\ 2007.\ Late formation and prolonged 
differentiation of the Moon inferred from W isotopes in lunar metals.\ 
{\it Nature} {\bf 450}, 1206-1209.

\item[] Thommes, E.W., Duncan, M.J., and Levison, H.F.:
2003, {\it Icarus} {\bf 161}, 431.

\item[] Thommes, E.W., and Duncan, M.J.: 2006, in {\it
Planet Formation: Theory, Observations, and Experiments}, ed.\
H.~Klahr and W.~Brander (Cambridge: Cambridge University Press), 129.

\item[] Ward, W.R.: 1986, {\it Icarus} {\bf 67}, 164.

\item[] Ward, W.R.: 1997, {\it Icarus} {\bf 126}, 261.

\item[] Weidenschilling, S.~J., Davis, D.~R.\ 1985.\ Orbital resonances in the 
solar nebula - Implications for planetary accretion.\ {\it Icarus} {\bf 62}, 16-29. 

\item[] Wetherill, G.W.~and Stewart, G.R.: 1989, {\it
Icarus} {\bf 77}, 330.

\item[] Wetherill, G.W.~and Stewart, G.R.: 1993, {\it
Icarus} {\bf 106}, 190.

\item[] Wisdom, J.~and Holman, M.: 1991, {\it Astronomical
Journal} {\bf 102}, 1528.

\item[] Wisdom, J.~and Holman, M.: 1991, {\it Astronomical
Journal} {\bf 102}, 1528.

\item[] Zappal{\`a}, V., Cellino, A., dell'Oro, A., Paolicchi, P.\ 
  2002.\ Physical and Dynamical Properties of Asteroid Families.\ 
  Asteroids III 619-631.

\end{itemize}

\clearpage

\end{document}